\documentclass[12pt]{article}
\pdfoutput=1
\usepackage{booktabs}
\usepackage{amsmath}
\usepackage{amssymb}
\usepackage{hhline}
\usepackage[scale={.75,.75}]{geometry}
\usepackage{url}
\usepackage{caption}
\usepackage[numbers, sort&compress]{natbib}
\usepackage{epsfig}
\usepackage{lscape}
\usepackage{mathrsfs}
\usepackage{multirow}
\usepackage{color}
\usepackage{verbatim}
\usepackage{nicefrac} 
\usepackage{upgreek}
 \usepackage{bbm}
\usepackage{psfrag}
\usepackage{hyperref}
 \usepackage{hyperref}

\newcommand{\q}{\textbf{q}}
\newcommand{\p}{\textbf{p}}
\newcommand{\tp}{(t_1+t_2)/2}
\newcommand{\tm}{t_1-t_2}
\renewcommand{\k}{\textbf{k}}

\newcommand{\Slash}[1]{\displaystyle{\not}{#1}}
\newcommand{\slashed}[1]{\displaystyle{\not}{#1}}

\newcommand{\fermionmass}[1]{\mathfrak{#1}}

\newcommand{\xx}{\rm x}

\newcommand{\pp}{{\rm p}}
\newcommand{\qq}{{\rm q}}
\newcommand{\kk}{{\rm k}}
\newcommand{\newalpha}{\rm a}
\newcommand{\newbeta}{\rm b}
\begin{document}
\date{\mbox{ }}

\title{ \vspace{-1cm}
%
\begin{flushright}
{\scriptsize \tt TUM-HEP-1020/15, NITS-PHY-2015008, CAS-KITPC/ITP-454}  
\end{flushright}
{\bf \boldmath 
 Sterile neutrino Dark Matter production from scalar decay in a thermal bath
 }
\\ [5mm]
}

\author{Marco Drewes$^{a, e}$, 
Jin U Kang$^{b, c, d, e}$ \\  \\
{\normalsize \it$^a$ Physik Department T70, Technische Universit\"at M\"unchen,} \\{\normalsize \it James Franck Stra\ss e 1, D-85748 Garching, Germany}\\
{\normalsize \it$^b$  Department of Physics,  Nanjing University,}\\
{\normalsize \it 22 Hankou Road, Nanjing, China 210093}\\
{\normalsize \it$^c$ Department of Physics,  Kim Il Sung University,}\\
{\normalsize \it RyongNam Dong, TaeSong District, Pyongyang, DPR Korea}\\
{\normalsize \it$^d$ International Centre for Theoretical Physics,}
{\normalsize \it Strada Costiera 11, 34151 Trieste, Italy}\\
{\normalsize \it $^e$ Kavli Institute for Theoretical Physics China, CAS, Beijing 100190, China\footnote{Visiting in September 2015.}}}

\maketitle

\thispagestyle{empty}

\begin{abstract}
  \noindent We calculate the production rate of singlet fermions from the decay of neutral or charged scalar fields in a hot plasma. We find that there are considerable thermal corrections when the temperature of the plasma exceeds the mass of the decaying scalar. We give analytic expressions for the temperature-corrected production rates in the regime where the decay products are relativistic. We also study the regime of non-relativistic decay products numerically. Our results can be used to determine the abundance and momentum distribution of Dark Matter particles produced in scalar decays. The inclusion of thermal corrections helps to improve predictions for the free streaming of the Dark Matter particles, which is crucial to test the compatibility of a given model with cosmic structure formation. With some modifications, our results may be generalised to the production of other Dark Matter candidates in scalar decays.
\end{abstract}
\newpage
\tableofcontents


\section{Introduction}\label{intro}
\subsection{Motivation} The motion of celestial bodies over a vast range of scales, ranging from sub-galactic to supercluster scales, deviates from the prediction of general relativity and the observed distribution of baryonic matter.
In addition, it seems impossible to explain the formation of the observed structures in the universe from the gravitational collapse of the primordial density perturbations observed in the CMB.
All attempts to consistently explain these phenomena on all scales by a modification of the laws of gravity have failed so far. 
All known data are, however,  in excellent agreement with the hypothesis that significant amounts of non-luminous \emph{Dark Matter} (DM) are present in the observable universe. 
This hypothesis is not only consistent with the observed movement of stars, galaxies and clusters as well as structure formation in the early universe (with the known laws of gravity), it also gives an explanation for the apparent gravitational lensing of light from distant galaxies and quasars.
If one takes this viewpoint, then these observations can be used to map the distribution of DM in the universe. The simplicity of this picture, the excellent agreement with data from very different physical environments and the lack of any known alternative explanation make the DM hypothesis very compelling. It is, however, not clear at this stage what the DM is made of.

The particles that compose the DM must be massive, electrically neutral, long lived and almost collisionless. The only candidates that fulfil these requirements within the SM are neutrinos.
The known neutrinos, however, have masses below 1 eV and were relativistic at the time when structures in the early universe formed. The free streaming of such \emph{Hot Dark Matter} (HDM) would suppress the formation of small scale structures in the universe, which is in contradiction with observations, see e.g. \cite{Frenk:2012ph} for a summary. Hence, DM must be made of one or more new particles. One obvious candidate would be heavier neutrino mass eigenstates, which could be consistent with structure formation constraints.
A comprehensive overview of particle physics scenarios that realise this idea and constraints on them from experiments, astrophysics and cosmology are given in \cite{Adhikari:2016bei}. In the following we recap only the basic facts that are relevant for the present work. 

\subsection{Heavy sterile neutrinos}  
Speculations about the existence of heavy neutrinos $N$ have been motivated for various other reasons, see \cite{Drewes:2013gca} for a review.
Adding new leptons with gauge charges to the SM requires to add a whole fourth generation of particles to ensure cancellations of all anomalies. This scenario is strongly constrained by experimental data, hence the heavy neutrinos should be ``sterile'', i.e., singlets under the SM gauge groups.
In the strict sense, we use the term \emph{sterile neutrinos} for singlet fermions that mix with the known SM neutrinos $\nu_L$, which are usually identified with right handed neutrinos $\nu_R$. 
However, many authors use the term in a more loose sense for any fermionic gauge singlet (or heavy neutral lepton), regardless of whether or not it mixes with $\nu_L$.
Since the scalar decay production mechanism we discuss here does not rely on a mixing of $N$ with ordinary neutrinos, our results apply in both cases.

\paragraph{Seesaw type heavy neutrinos}  - The probably strongest motivation for sterile neutrinos in the strict sense, i.e. right handed neutrinos $\nu_R$ that do mix with $\nu_L$, are the observed neutrino oscillations, which have been awarded with the 2015 Nobel Prize in physics. 
In the SM neutrinos only exist with left handed (LH) chirality, while all other fermions come with LH and right handed (RH) chirality. An explicit mass term for LH neutrinos $\nu_L$ is forbidden by gauge symmetry, and the Higgs mechanism cannot generate a Dirac mass term $\bar{\nu}_Lm_D\nu_R$ for them without existence of a RH counterpart $\nu_R$. If $\nu_R$ exist, they are allowed to have an explicit Majorana mass term $\bar{\nu}_RM_M\nu_R^c$ because they are gauge singlets. 
Here $\nu_R^c=C\overline{\nu_R}^T$.
If $\nu_R$ exists, the mixing between the known ``active'' (SU(2) charged) neutrinos $\nu_{L \alpha}$ (with $\alpha=e, \mu, \tau$) and the ``sterile'' singlets $\nu_{R I}$ (where $I=1\ldots n$ labels the families of RH neutrinos) generates a Majorana mass term for the $\nu_{L \alpha}$ and explains the neutrino oscillations via the seesaw mechanism \cite{Minkowski:1977sc,GellMann:seesaw,Mohapatra:1979ia,Yanagida:1980xy,Schechter:1980gr,Schechter:1981cv}.

The seesaw mechanism predicts the existence of $n$ new neutrino mass states $N_I$ that are almost identical with the singlets $\nu_{R I}$, but contain a small admixture $\theta_{\alpha I}\nu_{L \alpha}^c$ of the doublet fields. One can describe them by Majorana spinors 
\begin{eqnarray}\label{MassEigenstates}\label{Ndef}
N_I=\left[
V_N^\dagger\nu_R+\theta U_N^*\nu_{L}^{c} +\left(V_N^\dagger\nu_R+\theta U_N^*\nu_{L}^{c}\right)^c\right]_I.
\end{eqnarray}
Here $\theta=m_D M_M^{-1}$, $V_N$ is the equivalent of the Pontecorvo-Maki-Nakagawa-Sakata matrix in the sterile sector and $U_N$ is its unitary part. We can in good approximation take $V_N=U_N=1$.\footnote{We use the notation of \cite{Drewes:2015iva}, where all relations are defined more precisely.}
The active-sterile mixing $\theta$ leads to a $\theta$-suppressed weak interaction of the $N_I$, and this is the only way how they couple to the SM at low energies. There have been different suggestions what the role of the $N_I$ in cosmology and particle physics could be. Their CP-violating interactions could, for instance, generate the observed baryon asymmetry of the universe \cite{Canetti:2012zc} via leptogenesis during their decay \cite{Fukugita:1986hr} or production \cite{Akhmedov:1998qx,Asaka:2005pn} in the early universe. 
For very low eigenvalues $M_I$ of $M_M$, they could be responsible for the LSND \cite{Athanassopoulos:1996jb}, Gallium \cite{Abdurashitov:2005tb} and reactor \cite{Mention:2011rk} neutrino oscillation anomalies  and/or act as extra relativistic degrees of freedom in the early universe ("dark radiation"). 
Last but not least, the $N_I$ are obvious DM candidates: They inherit all advantages of the known neutrinos, but their masses are given by the eigenvalues of $M_M$ and can be much heavier.
Which of these roles the $N_I$ are able to take strongly depends on the magnitude of the $M_I$, see e.g.\ \cite{Drewes:2013gca,Drewes:2015jna}. For masses below the electroweak scale, various experimental constraints exist \cite{Hernandez:2014fha,Drewes:2015iva,Antusch:2015mia}, and the $N_I$ can be searched for in near future experiments \cite{Shrock:1980vy,Shrock:1981wq,Gorbunov:2007ak,Atre:2009rg,Alekhin:2015byh,Banerjee:2015gca,Blondel:2014bra,Antusch:2015mia,Canetti:2014dka,Mertens:2014nha,deGouvea:2015euy}.

\paragraph{Observational tests} - The idea that heavy neutrinos compose the DM can be tested in various different ways, leading to significant constraints on their properties. 
Let us in the following focus on one singlet fermion $N$ with mass $m_N$ that should compose the DM.
The most model independent constraint on its properties can be found by applying phase space considerations \cite{Tremaine:1979we} to DM dense regions, which imposes a lower bound on $m_N$.
The precise value of this bound depends on the $N$ phase space distribution function $f_N$, which depends on the way how $N$ were produced in the early universe.
It usually lies in the range of a few keV \cite{Gorbunov:2008ka}, $1-2$ keV can be used as a conservative estimate.

If $N$ is a sterile neutrino in the strict sense, we can identify
$N\equiv N_1$ in (\ref{Ndef}). Then $N$ at least interacts via its total mixing angle $U^2\equiv\sum_\alpha |\theta_{\alpha 1}|^2$.
This makes decays $N\rightarrow\nu\nu\nu$ into neutrinos possible, and the $N$-lifetime $\propto U^{-2}m_N^{-5}$ must be at least comparable to the age of the universe. Moreover, the radiative decay $N\rightarrow \gamma \nu$ \cite{Pal:1981rm,Barger:1995ty} predicts an emission line of energy $m_N/2$ from DM dense regions in space \cite{Abazajian:2001vt}.
Non-observation of this emission imposes an upper bound on $U^2$ for given $m_N$. 
In 2014, two independent publications reported the detection of an unidentified emission signal at 3.5 keV that could be interpreted in this way, though this claim is disputed \cite{Jeltema:2014qfa,Malyshev:2014xqa,Frandsen:2014lfa,Riemer-Sorensen:2014yda,Anderson:2014tza,Carlson:2014lla,Jeltema:2015mee,Ruchayskiy:2015onc}.  
For relatively small masses, the mixing $U^2$ can also lead to a signal in an upgraded version of the KATRIN experiment \cite{Mertens:2014nha,Barry:2014ika}. A direct detection \cite{Li:2010vy,Liao:2010yx}, on the other hand, would be very challenging \cite{Ando:2010ye}.

The free streaming of $N$ and its effect on structure formation provide another way to constrain the parameter space. These depend strongly on the $N$ phase space distribution function $f_N$, which is determined by the way the heavy neutrinos are produced in the early universe.
The precise shape of $f_N$ can be found by solving a set of momentum dependent kinetic equations for the coupled system composed of the scalar, $N$ and the thermal plasma, see e.g. \cite{Kaplinghat:2005sy,Kusenko:2006rh,Petraki:2007gq,Bezrukov:2014qda,Merle:2014xpa,Merle:2015oja}. A detailed discussion can be found in section 5 of \cite{Adhikari:2016bei}. 
Doing this requires knowledge of the $N$-production rate. 
In this work we calculate thermal corrections to the production rate due to interactions with the primordial plasma or a pre-existing $N$-population.

\paragraph{Sterile neutrino distribution function} - 
The framework of nonequilibrium quantum field theory allows to systematically compute thermal corrections to the production rate of heavy neutrinos.
Some elements of nonequilibrium quantum field theory are briefly summarised in the appendix \ref{methods}.
The field theoretical equivalent to the classical phase space distribution $f_{N\q}$ is given by the function $f_N(q_0)$ in the Kadanoff-Baym ansatz
(\ref{KBansatz})
for the heavy neutrino propagator when evaluated at the quasiparticle pole $\Upomega_{N \q}$,
\begin{equation}\label{PhaseSpaceDistrDef}
f_{N\q}=f_N(\Upomega_{N \q}).
\end{equation}
Due to the feeble interaction of the singlets $N$, we can for all practical purposes approximate $\Upomega_{N \q}\simeq\omega_{N\q}=(\q^2+m_N^2)^{1/2}$.
Our main results for the rate of $N$-production, which can be expressed in terms of one-dimensional integrals,  are valid for arbitrary $f_N$. That is important because $f_N$ is a dynamical quantity that changes throughout the production, and the production rate in each moment in time is affected by $f_N$ in that moment.

In addition to the general expressions, we provide illustrative analytic results by employing a simple parametrisation,\footnote{For $q_0<0$, $f_N(q_0)$ is determined by the relation (\ref{NegativefN}).}
\begin{equation}\label{fNAnsatz}
f_N(q_0)=\newalpha f_F(\newbeta q_0) \ {\rm for} \ q_0>0.
\end{equation}
The ansatz (\ref{fNAnsatz}) has a simple physical interpretation. $\newbeta=1$ corresponds to ``kinetic equilibrium'', i.e., the distribution function has the same shape as a Fermi-Dirac distribution, but a different normalisation. For $\newbeta\neq1$ the average momentum is shifted with respect to the background plasma temperature $T$: larger $\newbeta$ correspond to a smaller effective $N$-temperature. This could e.g. be realised if a $N$-population that was produced at earlier times has been diluted by entropy injection into the plasma after it decoupled.
One can constrain $\newalpha$ and $\newbeta$ by the requirement that $N$ with distribution $f_N$ at a given temperature $T$ make up for a fraction $r$ of the observed DM density $\Omega_{DM}$,
\begin{eqnarray}\label{DMfraction}
r\Omega_{DM}=\frac{m_N}{\rho_c}\frac{s_0}{s}\int \frac{d^3\p}{(2\pi)^3}f_N(\omega_{N\p})\simeq\frac{\newalpha}{\newbeta^3}\frac{8\pi m_N }{3 H_0^2 m_P^2}\frac{g_{s 0}}{g_s}2\frac{7}{8}\frac{\pi^2}{30}T_0^3,
\end{eqnarray}
where $s=2\pi^2g_s T^3/45$ is the radiation entropy density with the effective number $g_s$ of degrees of freedom, $m_P$ is the Planck mass and $H$ the Hubble parameter. All quantities with a subscript $_0$ refer to present day values.
Plugging in $\Omega_{DM}=0.268$ \cite{Ade:2013zuv} for $\newbeta=r=1$ yields $\newalpha\simeq 5\times 10^{-8} {\rm GeV}/m_M$, which implies $\newalpha<0.1$ for any $m_N$ that is consistent with phase space analysis bounds. 


\subsection{Singlet fermion production}

\paragraph{Thermal $N$ production via active-sterile mixing} - 
If $N$ is a sterile neutrino in the strict sense (\ref{Ndef}), then a minimal amount of $N$ is produced thermally via the mixing $\theta$ with active neutrinos \cite{Barbieri:1990vx} (Dodelson-Widrow mechanism \cite{Dodelson:1993je}). The expected free streaming classifies this contribution as Warm Dark Matter (WDM). 
The efficiency of the thermal production can be resonantly enhanced by the Mikheyev-Smirnov-Wolfenstein (MSW) effect if there are considerable lepton asymmetries $\mu_\alpha$ in the plasma \cite{Shi:1998km}.
The MSW resonance does not only affect the total number of produced $N$-particles, but also their momentum distribution $f_N$, which determines their free streaming length and thereby affects the formation of structures. 
The requirement to produce the right amount of DM from mixing alone imposes an upper and a lower bound on $U^2$ for given $m_N$. The upper bound, which requires the $N$-density to remain far below its equilibrium value, turns out to be weaker than the bounds from emission line searches and structure formation. The lower bound is crucial, but due to the MSW effect it depends on the value of the $\mu_\alpha$ during $N$-production. 

The requirement to explain the existence of small scale structures observed in Ly$\alpha$ absorption in quasar spectra puts an upper bound on the free streaming of DM. This already appears to exclude the scenario where all DM is composed of thermally produced $N$ for $\mu_\alpha=0$ (i.e. in absence of a MSW resonance) \cite{Horiuchi:2013noa,Merle:2014xpa}.
However, for $\mu_\alpha \neq 0$ the momentum distribution tends to be ``colder'' and can be consistent with structure formation \cite{Abazajian:2001nj,Boyarsky:2008xj,Lovell:2011rd,Kusenko:2009up} if $m_N>3.3$ keV \cite{Viel:2013apy}. 
It is difficult to work out all the details of the MSW-enhanced production because the $N$ production happens to peak at temperatures at which quarks start to form hadrons. Moreover, the MSW resonance produces non-thermal spectra, for which it is difficult to simulate structure formation \cite{Lovell:2011rd}.
In spite of significant progress \cite{Asaka:2006rw,Laine:2008pg,Ghiglieri:2015jua,Venumadhav:2015pla} considerable uncertainties remain.
At present it seems likely that thermally produced sterile neutrinos can explain all the observed DM consistent with the formation of small scale structure only if their production is resonantly enhanced by lepton asymmetries $\mu_\alpha$. 
The allowed  range for $m_N$ reaches from  $\sim 3.3$ keV to a few tens of keV, depending on $\mu_\alpha$. The mixing must be smaller than $U^2<10^{-8}- 10^{-12}$, depending on $m_N$.\footnote{One may wonder if such $N$ properties can arise in ``well-motivated'' theories of particle physics. An overview of possible models is e.g. given in \cite{Merle:2013gea}.}
Note that $N$ with such small mass and mixing can avoid constraints from the Cosmic Microwave Background or Big Bang Nucleosynthesis \cite{Hernandez:2014fha}.
However, they give no significant contribution to the generation of active neutrino masses in the seesaw mechanism. Hence, one needs at least two heavier siblings $N_2$ and $N_3$ to explain the two observed active mass splittings in the seesaw model. Interestingly, this minimal scenario is able to simultaneously explain neutrino oscillations, the observed DM density and the baryon asymmetry of the universe \cite{Asaka:2005pn,Canetti:2012vf,Canetti:2012kh}.
Alternatively, there can of course be an additional source of active neutrino masses. 

\paragraph{Production by gauge interactions} -
At low energies, $N$  in (\ref{MassEigenstates}) only interacts with other particles via its mixing $U^2$ with active neutrinos. This need not be true at high energies in the early universe, as $N$ are charged under some spontaneously broken gauge group in many extensions of the SM. In left-right symmetric theories, for instance, the $\nu_R$ (and hence $N$) are charged under a RH SU(2) gauge group. These interactions bring them into thermal equilibrium in the early universe, which would lead to a too large DM density.
There are different ways to avoid this problem.
The equilibration is avoided if the would-be freezeout temperature of the RH gauge bosons exceeds the (p)reheating temperature of the universe.  
This temperature is unknown; it may be constrained by CMB observations \cite{Martin:2010kz, Drewes:2015coa}.
If the $N$ equilibrate, their abundance  can be made consistent with observation if the number $g_*$ of degrees of freedom in the primordial plasma changes significantly after $N$-freezeout (e.g. because of a phase transition). Another (related) possibility is that the $N$ are diluted by a release of entropy into the primordial plasma after their freezeout \cite{Nemevsek:2012cd,Bezrukov:2009th,Bezrukov:2012as,Patwardhan:2015kga}.

\paragraph{$N$ production in scalar decays} -  
$N$-particles can also be produced non-thermally in the decay of a heavier particle.
This is the scenario under consideration here.
The heavy particle may be the inflaton \cite{Shaposhnikov:2006xi,Bezrukov:2009yw}, another scalar singlet \cite{Kusenko:2006rh,Petraki:2007gq,
Adulpravitchai:2014xna,Merle:2013wta,Kang:2014cia,Humbert:2015epa}, or a charged scalar \cite{Frigerio:2014ifa}.
Note that the minimal coupling to the SM already includes some production from the decays of scalars, namely pions \cite{Asaka:2006nq,Lello:2014yha,Lello:2015uma} and SM Higgs particles \cite{Bezrukov:2008ut,Matsui:2015maa}, which is, however, sub-dominant compared to the thermal production from mixing.
 Higgs decay may give a significant contribution in models with an additional leptophilic (or, more generally, DM-philic) Higgs doublet, which can decay into $N$ at much higher rate \cite{Haba:2014taa,Molinaro:2014lfa,Adulpravitchai:2015mna}.
The $N$ production may happen while the scalar field is in equilibrium (which is usually the case if it is charged) or while it ``freezes in'', leading to somewhat different phenomenology \cite{Merle:2015oja}. 
Since the decay mechanism does not rely on active-sterile mixing to produce the $N$, it does not impose a lower bound on $U^2$ and can avoid any constraints from searches for emission lines.
Moreover, it seems to lead to relatively cold DM spectra that are consistent with structure formation \cite{Merle:2014xpa}. 
In the present article, we compute thermal corrections to the abundance and momentum distribution of $N$-particles.

\subsection{Goal and outline of this work} 
In this paper we calculate thermal corrections to the production rate of sterile neutrinos $N$ in the decay of massive scalar fields $\phi$ (gauge singlet) and $\Phi$ (charged). 
For these considerations, it is irrelevant whether or not $N$ mixes with ordinary neutrinos; our results apply to any massive gauge singlet fermion that is produced in the decay of scalars.
If $N$ is a sterile neutrino in the sense of (\ref{Ndef}), 
then there is an additional population of $N$ that is produced thermally via the Dodelson-Widrow mechanism.
However, at the temperatures where scalar decays usually happen, this thermal production by mixing is negligible\footnote{
The production due to mixing  peaks at rather low temperatures $T\sim 100$ MeV.} and we can set $U^2=0$.
Then (\ref{MassEigenstates}) simplifies to $N\simeq \nu_R + \nu_R^c$, and $N$ approximately does not interact with the SM.

In the literature it is usually assumed that the production rate of $N$ particles with a given momentum $\p$ can be obtained from vacuum decay rate 
of scalar particles into heavy neutrinos. The momenta of the heavy neutrinos are uniquely fixed in this $1\rightarrow 2$ decay.  
However, in reality the decay happens in the hot and dense primordial plasma that filled the early universe, and it is well-known that the damping rate of scalars 
receives thermal corrections in this regime \cite{Weldon:1983jn}. There are several reasons for this. First, quantum statistics can enhance or suppress the decay rate, depending on whether the final state particles are bosons or fermions.
Second, the dispersion relations of quasiparticles in a dense plasma usually differ from those of particles in vacuum. This can change the phase space.
Third, at high temperature the decay $scalar \rightarrow N + something$ is not the only process that can produce $N$; inelastic scatterings and Landau damping can also contribute to the rate.

In this work we calculate thermal corrections to the DM production rate from first principles of nonequilibrium quantum field theory. This in principle is a very complicated calculation because the production of $N$ involves the scalar field, which is potentially far from thermal equilibrium during production (if it happens during freeze-in). To avoid calculations with nonequilibrium 
propagators of the scalar in the loop, we use a trick and calculate the thermal damping rate for the scalar-quasiparticles with momentum $\q$ due to the interactions with $N$. This can be done without specifying the thermodynamic state of the scalar field.
The total number of $N$-particles and their momentum distribution can be found by plugging the production rate into a set of Boltzmann equations for $N$ and the scalars.
Here we assume that the $N$-particles are produced in the decay of the scalar-particles. If they are produced in the decay of the condensate or ``classical field'' $\varphi=\langle\phi\rangle$ 
(or $\langle\Phi^\dagger\Phi\rangle^{1/2}$), the effective masses and couplings are generally $\varphi$-dependent. Thermal corrections to the dissipation rate in this case have e.g. been studied in \cite{Cheung:2015iqa} and references therein.

Compared to previous calculations, our method allows to include all quantum statistical factors and a proper treatment of the dispersion relations in the plasma.
It also systematically includes processes other than the decay to leading $\log$ accuracy in the gauge coupling. For instance, if $\Phi$ is electrically charged and decays into $N$ and a charged fermion $\Psi$ as $\Phi\rightarrow N\Psi$, then there are also scattering processes $\Psi\gamma \rightarrow\Phi N$ (s-channel) and $\gamma\Phi\rightarrow N \Psi$ (t-channel) as well as their inverse that change the number of $N$ and $\Phi$ particles, or the final state may radiate off a photon ($\Phi\rightarrow N\Psi \gamma$). Although these are of higher order in some coupling constant, they can become important at high density because of the Bose enhancement that photons are subject to.
The same happens if $\Phi$ is charged under a non-Abelian group and e.g.\ has electroweak interactions.
This has not been taken into account in past calculations.
In this work we calculate the leading order corrections to the production rate due to these effects.

The paper is organised as follows. In section \ref{SingletSec} we calculate the thermally corrected rate of $N$-particle production from neutral scalar decays.
In section \ref{ChargedSec} we calculate  the thermally corrected rate of $N$-particle production from U(1)-charged scalar decays.
The equivalent rate for SU(N) charged scalars can be obtained from that by a few simple replacements.
We also compute the leading log contribution to the $N$-production from scatterings in the plasma. 
The rates can in general only be determined numerically. We give analytic approximations that hold for the regime where the decay products are relativistic.
In section \ref{Conclusions} we discuss our results and conclude.
In the appendix \ref{methods} we briefly recall the methods from nonequilibrium quantum field theory required for this calculation, for more details see \cite{Berges:2004yj} (in general) and \cite{Anisimov:2008dz,Drewes:2010pf} (specifically for the approach used here).

\section{Scalar singlet decay}\label{SingletSec}
We first treat the case of production from the decay of a real singlet scalar field $\phi$.
That is, we consider the Lagrangian
\begin{eqnarray}
	\label{Lsinglet}
	\mathcal{L} &=&\mathcal{L}_{SM}+ 
	\frac{i}{2} \overline{N}\slashed{\partial}N-
	\overline{\ell_{L}}F N\tilde{h} -
	\tilde{h}^{\dagger}\overline{N}F^{\dagger}\ell_L - \frac{1}{2}\overline{N}m_NN\nonumber\\
&&+\frac{1}{2}\partial_\mu\phi\partial^\mu\phi - V(\phi) - \frac{1}{2} y\phi\overline{N}N
+\mathcal{L}_{\phi {\rm int}}.
		\end{eqnarray}
Here $h$ is the Higgs doublet and $\tilde{h}= \epsilon h^*$ with $\epsilon$ being the SU(2) antisymmetric tensor. $\ell_L$ is the SM lepton doublet, $F$ is a matrix of Yukawa couplings, the potential $V(\phi)$ includes a mass term $m_\phi^2\phi^2/2$ and $y$ is the Yukawa coupling that mediates the $\phi$ decay into $N$.  
Recall that $N=N^c$ is a Majorana spinor.
Since we are not interested in production via active sterile mixing, we immediately set $F=0$.
$\mathcal{L}_{\phi {\rm int}}$ contains interactions of $\phi$ with other fields than $N$. Such interactions must exist in order to produce $\phi$ in the early universe.
$\mathcal{L}_{\phi {\rm int}}$ may e.g.\ include a Higgs portal term $\tilde{\lambda}\phi^2h^\dagger h$, Yukawa interactions or other scalar interactions.

\subsection{Heavy neutrino production rate} 
In nonequilibrium and thermal field theory, gain- and loss rates for the occupation numbers of weakly coupled fields are related to the discontinuities of self-energies evaluated at the quasiparticle pole \cite{Weldon:1983jn}, see appendix \ref{methods}. 
Using the finite temperature optical theorem and cutting rules \cite{Kobes:1985kc}, one can interpret these rates in terms of decays and scatterings involving the external particles as well as the ``cut'' propagators in the loop \cite{Bedaque:1996af,Landshoff:1996ta}.
In principle, the most straightforward way to calculate the number and momentum distribution of $N$-particles of a given momentum $\p$ that are produced in a given time interval is to compute the $N$-self energies $\Sigma^\gtrless_N$ in the closed time path formalism and obtain rates of the form (\ref{FermionGamma}). 
This, however, requires evaluating Feynman diagrams with nonequilibrium propagators for both, the decaying scalar $\phi$ and $N$, in the loop. The interactions of the singlet fields $N$ are typically feeble, which justifies to use the Kadanoff-Baym ansatz (\ref{KBansatz}) and apply the narrow width approximation ($\Upgamma_{N\q}=0$) for propagators inside the loops. 
For the scalar field it is not obvious that these approximations are always justified because it usually has other interactions in addition to the coupling to $N$. 
In the present work we are interested in scenarios where $N$-production is dominated by decays.
In this regime we can use a trick to avoid the scalar propagator in the loops: Instead of directly calculating the production rate of $N$, we compute the damping rate of $\phi$ due to interactions with $N$ and use it to calculate the number of produced $N$-particles via (\ref{momdepNprod}).

\paragraph{Kinetic equations for the occupation numbers} - 
The occupation numbers for $\phi$-modes can be characterised by a function $f_{\phi\q}(t)$, which follows the effective kinetic equation (\ref{EffBE}),
\begin{eqnarray} \nonumber
  \partial_{t} f_{\phi\q}  &=& \left[ 1+f_{\phi\q}  \right]\Gamma_{\phi\q}^<  - f_{\phi\q} \Gamma_{\phi\q}^>  \nonumber    \\
&=&-\Gamma_{\phi\q} \left[f_{\phi\q} -\bar f_{\phi\q} \right] \label{KineticEQ}
\end{eqnarray}
with 
\begin{equation}
\bar f_{\phi\q} \equiv(\Gamma_{\phi\q}^> /\Gamma_{\phi\q}^< -1)^{-1}.
\end{equation}
Here $\Gamma_{\phi\q}^\gtrless$ are gain- and loss rates and $\Gamma_{\phi\q}=\Gamma_{\phi\q}^>-\Gamma_{\phi\q}^<$ can be identified with the total thermal damping rate.
The precise definitions of these rates are given in the appendix by (\ref{gain}), (\ref{loss}) and (\ref{GammaFormula2}).
In the following we will use the symbol $\tilde{\Gamma}_{\phi\q}$ for the contribution to the thermal $\phi$-width $\Gamma_{\phi \q}$
that comes from diagrams with $N$-propagators in the loop.
We can formally define $\tilde{\Gamma}_{\phi\q}$ by the equation
\begin{equation}
\Gamma_{\phi\q} \equiv \tilde{\Gamma}_{\phi\q} + \Gamma_{\phi\q}^{(0)},
\end{equation}
where $\Gamma_{\phi \q}$ is the total thermal $\phi$-width and
$\Gamma_{\phi\q}^{(0)}\equiv \Gamma_{\phi \q}|_{y=0}$ is $\Gamma_{\phi \q}$ at zeroth order in $y$. $\Gamma_{\phi\q}^{(0)}$ is generated from the interactions in $V(\phi)$ and $\mathcal{L}_{\phi {\rm int}}$. Analogously, one can define self energies $\tilde{\Pi}_{\phi}^\gtrless$ and rates $\tilde{\Gamma}_{\phi\q}^\gtrless$ from (\ref{gain}) and (\ref{loss}).
For convenience, we can rewrite \eqref{KineticEQ} as 
\begin{eqnarray}\label{convenientlyrewritten}
\partial_t f_{\phi\q}  = -[\tilde{\Gamma}_{\phi\q} +\Gamma_{\phi\q}^{(0)} ]\left[f_{\phi\q} -\bar f_{\phi\q} \right].
\end{eqnarray}
Note that $\bar{f}_{\phi\q} $ implicitly depends on $y$ and $f_{N}$ because it has to be evaluated with the full $\Gamma_{\phi\q}^\gtrless$. 
If the interactions in $\mathcal{L}_{\phi {\rm int}}$ are numerous or much stronger than the Yukawa coupling (e.g. $\tilde{\lambda}\gg y$), then one may expand in $\tilde{\Gamma}_{\phi\q}/\Gamma_{\phi\q}^{(0)} \ll 1$ to obtain 
\begin{eqnarray}
\partial_t f_{\phi\q}  \simeq -[\tilde{\Gamma}_{\phi\q} +\Gamma_{\phi\q}^{(0)} ]\left[f_{\phi\q} -\bar f_{\phi\q}^{(0)} \right],
\end{eqnarray}
where $f_{\phi\q}^{(0)} \equiv f_{\phi\q} |_{y=0}$ is independent of $y$ and $f_N$.
The total number of $N$-particles 
\begin{equation}
n_N \equiv \int\frac{d^3\p}{(2\pi)^3}f_{N \p} 
\end{equation}
is given by the rate equation
\begin{eqnarray}\label{RateEquation}
\partial_t n_N  = 2\int\frac{d^3\q}{(2\pi)^3} \tilde{\Gamma}_{\phi\q} \left[f_{\phi\q} -\bar f_{\phi\q} \right].
\end{eqnarray}
The factor 2 in front of the integral is due to the fact that two $N$-particles are produced in each $\phi$-decay.
If the main contribution to $\tilde{\Gamma}_{\phi \q}$ comes from $1\rightarrow 2$ decays and their inverse, 
one can express
\begin{eqnarray}\label{DefOfgamma}
\tilde{\Gamma}_{\phi\q}=\int\frac{d^3\p}{(2\pi)^3}\tilde{\gamma}_{\phi}(\p,\q) \delta(\Omega_{\phi \q} - \Upomega_{N \p} - \Upomega_{N \p-\q}).
\end{eqnarray}
and
\begin{eqnarray}\label{momdepNprod}
\partial_t f_{N\p} &=& \int\frac{d^3\q}{(2\pi)^3}
[\tilde{\gamma}_{\phi}(\p,\q) + \tilde{\gamma}_{\phi}(\q-\p,\q)]
\left[f_{\phi\q}-\bar f_{\phi\q} \right]\delta(\Omega_{\phi \q} - \Upomega_{N \p} - \Upomega_{N \p-\q})\nonumber\\
&=& 2\int\frac{d^3\q}{(2\pi)^3}
\tilde{\gamma}_{\phi}(\p,\q)\left[f_{\phi\q}-\bar f_{\phi\q} \right]\delta(\Omega_{\phi \q} - \Upomega_{N \p} - \Upomega_{N \p-\q}).  
\end{eqnarray}
Here $\Omega_{\phi \q}$ is the mass shell of a $\phi$-quasiparticle with spatial momentum $\q$, see (\ref{OmegaAndGammaDefs}), and $\Upomega_{N \p}$ the energy for a $N$-quasiparticle with spatial momentum $\p$. In the second equality we used $\tilde{\gamma}_{\phi}(\p,\q) = \tilde{\gamma}_{\phi}(\q-\p,\q)$ for $\Omega_{\phi \q} = \Upomega_{N \p} + \Upomega_{N \p-\q}$. 
We should recall that the quantities $\Omega_{\phi \q}$, $f_{\phi\q}$, $\bar f_{\phi\q} $, $\tilde{\gamma}_{\phi}(\p,\q)$ and $f_{N\p}$ all depend on time.
In the following, we focus on the computation of the quantities  $\tilde{\Gamma}_{\phi\q}^>$ and $\tilde{\Gamma}_{\phi\q}^<$ and $\tilde{\gamma}_{\phi\q}$, i.e., the gain- and loss rates for $N$-production. From (\ref{convenientlyrewritten})-(\ref{momdepNprod}) it is clear that the full $\Gamma_{\phi}^{\gtrless}$ is needed to determine the time evolution of $f_{\phi\q}$ and calculate $f_{N\p}$ in general.

In the special case that $\phi$ predominantly couples to fields that are in equilibrium, $\Gamma_{\phi\q}^\gtrless$ fulfil a relation of the type (\ref{DetailedBalance}) and one can further approximate
\begin{eqnarray}
\partial_t f_{N\p} = 2\int\frac{d^3\q}{(2\pi)^3}
\tilde{\gamma}_{\phi}(\p,\q)\left[f_{\phi\q}-f_B(\Omega_{\phi\q})\right]\delta(\Omega_{\phi \q} - \Upomega_{N \p} - \Upomega_{N \p-\q}), \label{KEQ}
\end{eqnarray}
where $f_B(\Omega)= (e^{\Omega/T}-1)^{-1}$ is the Bose-Einstein distribution and we have suppressed the time dependence of all quantities. 
No knowledge of the full $\Gamma_{\phi \q}$ is needed to compute $f_B(\Omega_{\phi\q})$. 
However, $\Omega_{\phi\q}$ depends on time, and the parameters in $\mathcal{L}_{\phi {\rm int}}$ still affect $f_{N\p}$ because the time evolution of $f_{\phi\q} $ is governed by (\ref{convenientlyrewritten}).

\paragraph{The damping rate} -
As summarised in the appendix, $\Gamma_{\phi\q}$ can be determined from (\ref{GammaFormula}) by calculating the imaginary part of the retarded self-energy for $\phi$ at the quasiparticle pole,
 \begin{equation}
\Gamma_{\phi\q}\simeq-\frac{{\rm Im}\Pi_\phi^R(q)}{q_0}\Big|_{q_0=\Omega_{\phi \q}}.
\end{equation}
$\tilde{\Gamma}_{\phi\q}$ is obtained from the $y$-dependent part of the self-energy ${\rm Im}\tilde{\Pi}_\phi^R$.
The rates $\tilde{\Gamma}_{\phi\q}^\gtrless$ are obtained equivalently from $\tilde{\Pi}_\phi^\gtrless$.
At one-loop level, the self energies $\tilde{\Pi}_\phi^\gtrless$ can be calculated from the diagram
\begin{center}
\includegraphics[width=0.3\textwidth]{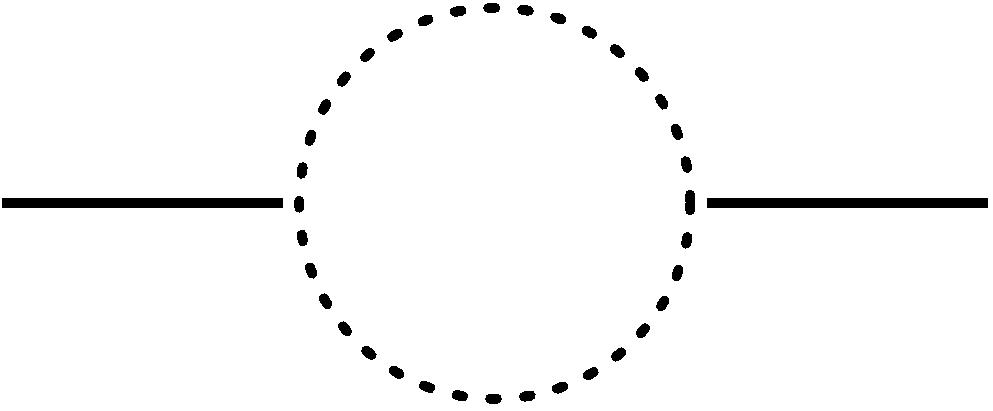}
\end{center}
by applying finite temperature Feynman rules \cite{LeB}.
Here the solid lines represent the external $\phi$ and the dotted lines $N$-propagators. 
Using the expressions given in appendix \ref{methods}, 
the self-energies read
\begin{eqnarray}
\tilde{\Pi}_\phi^<(q)&=&\frac{iy^2}{2}\int\frac{d^4 p}{(2\pi)^4}{\rm tr}\left[S_N^<(p)S_N^>(p-q)\right]\nonumber\\
&=&-\frac{iy^2}{2}\int\frac{d^4 p}{(2\pi)^4}\left(1-f_N(p_0-q_0)\right)f_N(p_0) \label{singletproduction<}
{\rm tr}\left[
\uprho_N(p)\uprho_N(p-q)
\right]
\end{eqnarray}
and
\begin{eqnarray}
\tilde{\Pi}_\phi^>(q)&=&\frac{iy^2}{2}\int\frac{d^4 p}{(2\pi)^4}{\rm tr}\left[S_N^>(p)S_N^<(p-q)\right]\nonumber\\
&=&-\frac{iy^2}{2}\int\frac{d^4 p}{(2\pi)^4}\left(1-f_N(p_0)\right)f_N(p_0-q_0)
{\rm tr}\left[
\uprho_N(p)\uprho_N(p-q)
\right].
\end{eqnarray}
In combination, this gives
\begin{eqnarray}\label{singletproduction}
\tilde{\Pi}_\phi^-(q)=\frac{iy^2}{2}\int\frac{d^4 p}{(2\pi)^4}\left(f_N(p_0)-f_N(p_0-q_0)\right){\rm tr}\left[
\uprho_N(p)\uprho_N(p-q)
\right].
\end{eqnarray}
If $N$ had gauge interactions while being produced in the primordial plasma, then this would not only add another efficient $N$-production mechanism (thermal production via gauge interactions), but would also modify the production from scalar decays \cite{Weldon:1983jn,Keil:1988sp}. 
In this case the results from example IV in section 7 of \cite{Drewes:2013iaa} may be applied. We will not repeat this calculation here.
Instead we assume that $N$ is indeed a singlet not only under the SM gauge groups, but with respect to all interactions that are relevant in the early universe.

\subsection{Small Yukawa couplings $y$: Production in decays}  
It is often assumed that the coupling $y$ is very small, $y<10^{-5}-10^{-9}$ \cite{Merle:2015oja}.
Then the term $y\phi\overline{N}N$ can be used to generate a keV scale Majorana mass term from a GeV-TeV vev of $\phi$, without having to add a small Majorana mass $\overline{N}m_N N/2$ in (\ref{Lsinglet}) by hand.
In this case the effects of forward scatterings (which modify the $N$ dispersion relation in the plasma) and inelastic scatterings (which increase the thermal width of $N$ and thereby reduce its effective lifetime) are negligible even if there is already some $N$-population in the plasma.
Then we can approximate $\uprho_N(p)$ with the free spectral density,
\begin{equation}\label{rhoNfree}
\uprho_N(p)\simeq 2\pi {\rm sign}(p_0)(\slashed{p}+m_N)\delta(p^2-m_N^2). 
\end{equation}
This essentially means that we use free thermal propagators for $N$ (instead of resummed propagators), which greatly simplifies the evaluation of the integrals in  \eqref{singletproduction<}-(\ref{singletproduction}). 
It also implies that the energy for $N$-quasiparticles in the plasma is in good approximation
\begin{equation} 
\Upomega_{N \p}\simeq\omega_{N \p}=(\p^2+m_N^2)^{1/2}.
\end{equation}
Corrections are suppressed by both, the smallness of $y\ll1$ and $f_N\ll 1$ due to (\ref{DMfraction}).

\paragraph{Calculation of the self-energies} -
We first focus on the case $\q=0$.
The results are
\begin{eqnarray}
\tilde{\Pi}_\phi^>(q)\big|_{\q=0}=  \frac{-iy^2}{8\pi}\,q_0^2\,\left[
1-\left(\frac{2m_N}{q_0}\right)^2
\right]^{3/2}
\left[1- f_N(q_0/2)\right]^2 \theta(q_0-2m_N),
\end{eqnarray}
\begin{eqnarray}
\tilde{\Pi}_\phi^<(q)\big|_{\q=0}=  \frac{-iy^2}{8\pi}\,q_0^2\,\left[
1-\left(\frac{2m_N}{q_0}\right)^2
\right]^{3/2}
\left[f_N(q_0/2)\right]^2\,  \theta(q_0-2m_N),
\end{eqnarray} 
hence
\begin{eqnarray}\label{singletzeromode}
\tilde{\Pi}_\phi^-(q)\big|_{\q=0}=  \frac{-iy^2}{8\pi}\,q_0^2\,\left[
1-\left(\frac{2m_N}{q_0}\right)^2
\right]^{3/2}
\left[1-2f_N(q_0/2)\right]\theta(q_0-2m_N).
\end{eqnarray}


Let us now turn to the case $\q\neq0$. In a thermal bath, $\Gamma_{\phi\q}$ in general cannot be obtained by multiplying $\tilde{\Gamma}_{\phi \q=0}$ with the time dilatation factor $m_\phi/(m_\phi^2+\q^2)^{1/2}$. 
This can be understood by recalling that $1/\Gamma_{\phi\q}$ is the lifetime of a quasiparticle and determines its mean free path in the bath. The difference in the lifetime of a $\q=0$ and $\q\neq0$ quasiparticle in a bath is not only due to time dilatation, but also due to the fact that the cross section for scatterings with bath particles (which reduces the lifetime) generally depends on the energy of the colliding particles. Moreover, the quantum statistical factor $(1-2f_N)$ breaks Lorentz invariance: A moving $\phi$-particle decays into $N$ with different momenta than one at rest, and the occupation numbers in those modes generally differ.
In the approximation (\ref{rhoNfree}) for very small $y$, the effect of scatterings is negligible, but quantum statistics may still play a role.
To evaluate the integrals in \eqref{singletproduction<}-(\ref{singletproduction}) at $\q\neq0$, we neglect the mass $m_N$ for simplicity, which seems justified for $m_N\sim$ keV.
Using $\delta(p_0^2)$ in $\uprho_N(p)$, it is straightforward to obtain
\begin{eqnarray}
\lefteqn{\tilde{\Pi}_\phi^>(q)=\frac{-iy^2}{2\pi}\int_0^\infty d\pp \int_{-1}^1 \pp\, d\xx \, \left[1-f_N(\pp)\right] f_N(\pp-q_0)}\nonumber\\
&\times&{\rm sign}(\pp -q_0)\, (\pp\qq \xx - \pp q_0)\, \delta[q_0^2-\qq^2+2\pp(\qq \xx -q_0)] 
- \big[q_0 \rightarrow -q_0 \big], 
\end{eqnarray}
\begin{eqnarray}
\lefteqn{\tilde{\Pi}_\phi^<(q)=\frac{-iy^2}{2\pi}\int_0^\infty d\pp \int_{-1}^1 \pp\, d\xx\, f_N(\pp)\left[1 -f_N(\pp-q_0)\right]}\nonumber\\
&\times&{\rm sign}(\pp -q_0)\, (\pp\qq \xx - \pp q_0)\, \delta[q_0^2-\qq^2+2\pp(\qq \xx -q_0)] 
- \big[q_0 \rightarrow -q_0 \big], 
\end{eqnarray}
\begin{eqnarray}
\lefteqn{\tilde{\Pi}_\phi^-(q)=\frac{iy^2}{2 \pi}\int_0^\infty d\pp \int_{-1}^1 \pp\, d\xx \left[f_N(\pp) -f_N(\pp-q_0)\right]}\nonumber\\
&\times&{\rm sign}(\pp -q_0)\, (\pp\qq \xx - \pp q_0)\, \delta[q_0^2-\qq^2+2\pp(\qq \xx -q_0)] 
- \big[q_0 \rightarrow -q_0 \big], 
\end{eqnarray}
where $\pp=|\p|$, $\qq=|\q|$  and $\xx=\p\q/\pp\qq$ is the cosine of the angle between the spatial vectors $\p$ and $\q$. 
The remaining $\delta$-function can be used to evaluate the $\xx$-integral, which fixes the limits for the $\pp$-integration:
\begin{eqnarray} \label{singletnonyeroresultArbitraryfN>}
\tilde{\Pi}_\phi^>(q)=\frac{-iy^2}{8\pi\qq}\int_{(q_0-\qq)/2}^{(q_0+\qq)/2}d\pp \, \left[1-f_N(\pp)\right]\left[1- f_N(q_0-\pp)\right]\,(q_0^2-\qq^2),
\end{eqnarray}
\begin{eqnarray} \label{singletnonyeroresultArbitraryfN<}
\tilde{\Pi}_\phi^<(q)=\frac{-iy^2}{8\pi\qq}\int_{(q_0-\qq)/2}^{(q_0+\qq)/2}d\pp  \, f_N(\pp) f_N(q_0-\pp) \,(q_0^2-\qq^2),
\end{eqnarray}
\begin{eqnarray}\label{singletnonyeroresultArbitraryfN}
\tilde{\Pi}_\phi^-(q)=\frac{-iy^2}{8\pi\qq}\int_{(q_0-\qq)/2}^{(q_0+\qq)/2}d\pp \left(1-f_N(\pp)-f_N(q_0-\pp)\right)(q_0^2-\qq^2).
\end{eqnarray}
It is worth to note that no assumptions about $f_N$ were required to obtain these expressions. 
If $f_N$ can be approximated by (\ref{fNAnsatz}), the final integrals can be solved analytically, e.g.\ 
\begin{eqnarray}\label{singletnonyeroresult}
\tilde{\Pi}_\phi^-(q)=\frac{-iy^2}{8\pi}
(q_0^2-\qq^2)
\left[
1-2\newalpha\left(
1
-\frac{T}{\newbeta \qq}{\rm log}\left[
\frac{
f_F(\newbeta (q_0-\qq)/2)
}{
f_F(\newbeta (q_0+\qq)/2)
}
\right]
\right)
\right].
\end{eqnarray}
If there is no pre-existing $N$-population present at the onset of the $\phi$-decay, then we can set $f_N=0$ and all temperature dependent effects in (\ref{singletnonyeroresult}) seem to disappear. This is what one would intuitively expect, and this is why the vacuum decay rate has been used in the past literature. 

\paragraph{Production rate} -
From (\ref{GammaFormula2}) it is clear that $\tilde{\Pi}_\phi^-$ should be evaluated at the $\phi$-quasiparticle pole $\Omega_{\phi \q}$. The $\phi$-quasiparticle dispersion relation  generally differs from 
\begin{equation}
\omega_{\phi\q}=(m_\phi^2+\q^2)^{1/2}, 
\end{equation}
that is, $\Omega_{\phi \q}\neq \omega_{\phi\q}$. This is true even if $\phi$ is a singlet because $\phi$ can have Yukawa couplings or couplings to other scalars, and $V(\phi)$ in general contains self-interactions. 
In order to be produced in the early universe, $\phi$ necessarily must have some interactions, and these will necessarily shift the quasiparticle mass pole.
If, for example
\begin{equation}\label{ScalarPotential}
V(\phi)=\frac{1}{2}m_\phi^2\phi^2 + \frac{\lambda}{4!}\phi^4,
\end{equation}
then 
\begin{equation}\label{quasiparticleenergyapprox}
\Omega_{\phi \q}^2\simeq\q^2+M_\phi^2, 
\end{equation}
with the thermal mass
\begin{equation}\label{ThermalMass}
M_\phi^2 = m_\phi^2 + \frac{\lambda}{24}T^2.
\end{equation}
Hence, even for $f_N=0$ there is a thermal correction to $\tilde{\Gamma}_{\phi\q}$.
To estimate this effect, we  compare (\ref{singletzeromode}) to the 
vacuum decay rate ($m_\phi \gg m_N$)
\begin{equation}\label{Gammaphi0Def}
\tilde{\Gamma}_0=\frac{y^2}{16\pi}m_\phi
\end{equation}
and express\footnote{Note that (\ref{Gammasingletzeromode21})
and (\ref{Gammasingletzeromode12}) fulfil a modified detailed balance relation even though $N$ and $\phi$ are not in equilibrium.
}
\begin{eqnarray}\label{Gammasingletzeromode21}
\tilde{\Gamma}_{\phi \q}^>|_{\q=0}=
\tilde{\Gamma}_0\frac{M_\phi}{m_\phi}
\left[
1-\left(\frac{2m_N}{M_\phi}\right)^2
\right]^{1/2}
\left[1-f_N(M_\phi/2)\right]^2,
\end{eqnarray}
\begin{eqnarray}\label{Gammasingletzeromode12}
\tilde{\Gamma}_{\phi \q}^<|_{\q=0} =
\tilde{\Gamma}_0\frac{M_\phi}{m_\phi}
\left[
1-\left(\frac{2m_N}{M_\phi}\right)^2
\right]^{1/2}
f_N(M_\phi/2) ^2,
\end{eqnarray}
\begin{eqnarray}\label{Gammasingletzeromode}
\tilde{\Gamma}_{\phi \q}|_{\q=0}=
\tilde{\Gamma}_0\frac{M_\phi}{m_\phi}
\left[
1-\left(\frac{2m_N}{M_\phi}\right)^2
\right]^{1/2}
\left[1-2f_N(M_\phi/2)\right]
.\end{eqnarray}
This can be compared to the production rate of a Dirac fermion at temperatures below its mass in section 7.1 of \cite{Drewes:2013iaa}. The only difference is an overall factor 2 for the Dirac fermion. 
This difference arises from the fact that, for a Dirac fermion, there are two diagrams of the type considered here, with a fermion flow in the loop going in opposite directions. More physically, a Dirac fermion has twice more internal degrees of freedom into which $\phi$ can decay. 
If one would calculate corrections to (\ref{singletzeromode}) by using dressed spectral densities instead of (\ref{rhoNfree}), then one would generally see further differences between the production of Dirac and Majorana fermions because the interactions of Majorana fermions are generally different. However, for small $y$ such corrections are negligible, and the free spectral density for Dirac and Majorana fermions is the same, hence the factor 2 is the only difference.
It is worth to emphasise that the expression (\ref{Gammasingletzeromode}) holds for $\emph{any}$ sterile neutrino distribution function $f_N$, i.e., it does not rely on the ansatz (\ref{fNAnsatz}). 

\begin{figure}
	\center
	\includegraphics[width=0.7\textwidth]{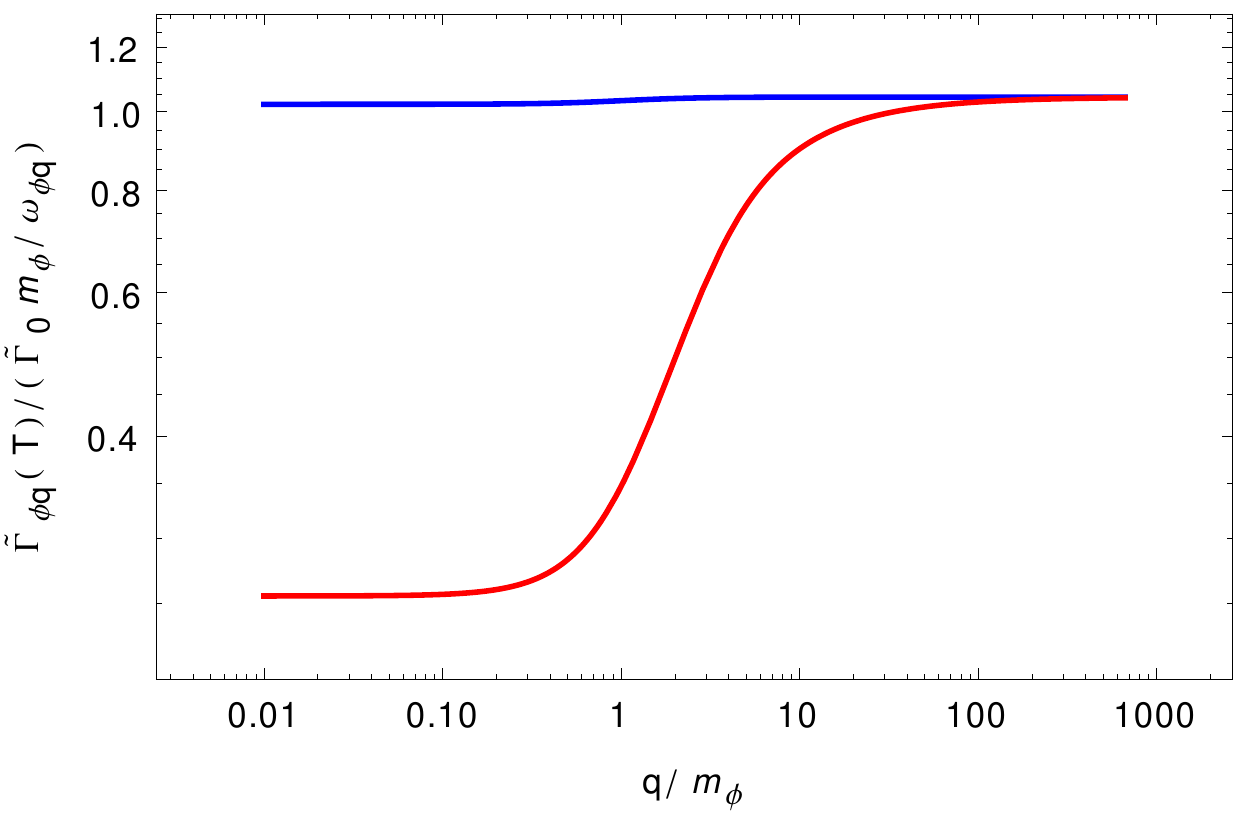}\\
\includegraphics[width=0.7\textwidth]{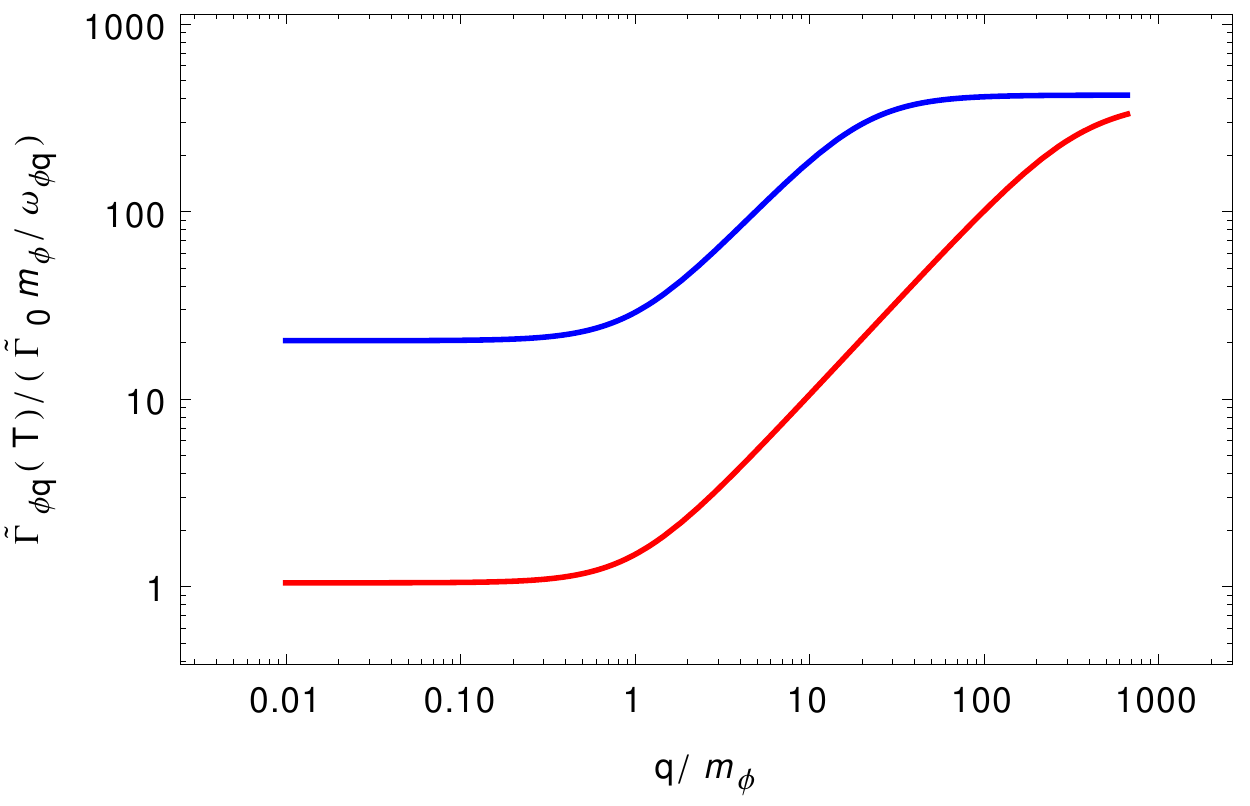}
\caption{The rate $\tilde{\Gamma}_{\phi\q}$ in units of the time-dilated rate $\tilde{\Gamma}_0 m_\phi/\omega_{\phi\q}$ with $M_\phi$ given by (\ref{ThermalMass}) and $\lambda=1$ as a function of $\qq$. The blue curve is for $\newalpha=0$, the red curve for $\newalpha=1$.
In the upper plot $T=m_\phi$, in the lower plot $T=100m_\phi$.
At high temperatures, the increased thermal mass $M_\phi$ leads to a larger rate than the vacuum estimate $\tilde{\Gamma}_0 m_\phi/\omega_{\phi\q}$ suggests. 
The Pauli blocking for $\newalpha\neq0$ is inefficient for momenta $\qq\gg T$ because the produced particles' momenta are outside the Fermi sphere.
\label{SingletqT}}
\end{figure}

Based on \eqref{singletnonyeroresultArbitraryfN>}-(\ref{singletnonyeroresultArbitraryfN}), one can calculate rates 
for arbitrary  $\qq\neq0$ as
\begin{eqnarray} \label{singletGammaArbitraryfN>}
\tilde{\Gamma}_{\phi \q}^>=\frac{y^2}{16\pi}\frac{\Omega_{\phi \q}^2-\qq^2}{\Omega_{\phi \q} \qq}\int_{(\Omega_{\phi \q}-\qq)/2}^{(\Omega_{\phi \q}+\qq)/2}d\pp \, \left[1-f_N(\pp)\right]\left[1- f_N(\Omega_{\phi \q}-\pp)\right],
\end{eqnarray}
\begin{eqnarray} \label{singletGammaArbitraryfN<}
\tilde{\Gamma}_{\phi \q}^<=\frac{y^2}{16\pi}\frac{\Omega_{\phi \q}^2-\qq^2}{\Omega_{\phi \q} \qq}\int_{(\Omega_{\phi \q}-\qq)/2}^{(\Omega_{\phi \q}+\qq)/2}d\pp  \, f_N(\pp) f_N(\Omega_{\phi \q}-\pp),
\end{eqnarray}
\begin{eqnarray}\label{singletGammaArbitraryfN}
\tilde{\Gamma}_{\phi\q}=\frac{y^2}{16\pi}\frac{\Omega_{\phi \q}^2-\qq^2}{\Omega_{\phi \q} \qq}\int_{(\Omega_{\phi \q}-\qq)/2}^{(\Omega_{\phi \q}+\qq)/2}d\pp \left[1-f_N(\pp)-f_N(\Omega_{\phi \q}-\pp)\right].
\end{eqnarray}
Here we have neglected the temperature dependence of $\mathcal{Z}$.\footnote{This is justified because the leading order thermal correction to ${\rm Re}\tilde{\Pi}_\phi^R(p)$ in (\ref{residue}) is usually momentum independent, c.f. (\ref{ThermalMass}), hence the deviation of $\mathcal{Z}$ from unity is of higher order. }
With the approximation (\ref{quasiparticleenergyapprox}), one finds
\begin{eqnarray}
\tilde{\Gamma}_{\phi \q}^>=\frac{y^2}{16\pi}\frac{M_\phi}{\Omega_{\phi \q}}\frac{M_\phi}{\qq}\int_{(\Omega_{\phi \q}-\qq)/2}^{(\Omega_{\phi \q}+\qq)/2}d\pp \, \left[1-f_N(\pp)\right]\left[1- f_N(\Omega_{\phi \q}-\pp)\right]
\end{eqnarray}
and
\begin{eqnarray}
\tilde{\Gamma}_{\phi \q}^<=\frac{y^2}{16\pi}\frac{M_\phi}{\Omega_{\phi \q}}\frac{M_\phi}{\qq}\int_{(\Omega_{\phi \q}-\qq)/2}^{(\Omega_{\phi \q}+\qq)/2}d\pp  \, f_N(\pp) f_N(\Omega_{\phi \q}-\pp).
\end{eqnarray}
One can combine them to express $\tilde{\Gamma}_{\phi\q}$ in terms of the vacuum decay rate for particles at rest $\tilde{\Gamma}_0$,
\begin{eqnarray}\label{singletGammaArbitraryfN2}
\tilde{\Gamma}_{\phi\q}=
\tilde{\Gamma}_0
\frac{M_\phi}{m_\phi}
\frac{M_\phi}{\Omega_{\phi \q}}
\frac{1}{\qq}\int_{(\Omega_{\phi \q}-\qq)/2}^{(\Omega_{\phi \q}+\qq)/2}d\pp \left[1-f_N(\pp)-f_N(\Omega_{\phi \q}-\pp)\right]
\end{eqnarray}
The physical interpretation of the different factors in this expression is very simple. $\tilde{\Gamma}_0$ is simply the decay rate for $\phi$-particles at rest in vacuum. 
The factor $M_\phi/m_\phi$ takes account of the fact that the effective mass of $\phi$-quasiparticles in the plasma is larger than in vacuum, leading to a reduced lifetime. The factor $M_\phi/\Omega_{\phi \q}$ takes account of the fact that the lifetime of a particle that moves with momentum $\q$ is larger than that of a particle at rest due to time dilatation. In vacuum, the dilatation factor would be given by  $m_\phi/\omega_{\phi \q}$; this cannot be applied in the thermal plasma (c.f.\ figure \ref{SingletqT}), but the obvious generalisation $M_\phi/\Omega_{\phi \q}$ works. What is left is a quantum statistical factor that breaks Lorentz invariance (because the thermal bath singles out a reference frame with respect to which $\q$ is measured). For $\q=0$ it reduces to the usual expression $1-2f_N(M_\phi/2)$.
From \eqref{DefOfgamma} we find
\begin{eqnarray}
\tilde{\gamma}_{\phi}(\p,\q) &=& \frac{-y^2\pi}{2\Omega_{\phi \q}\omega_{N \p}\omega_{N \p-\q}}
(\Omega_{\phi\q}\omega_{N \p} -2m_N^2-\p\q)
[1-f_N(\omega_{N \p})-f_N(\omega_{N \p-\q})] \nonumber \\
&\simeq&\frac{y^2\pi}{2\Omega_{\phi \q} |\p-\q|}
(\Omega_{\phi\q}-\qq\xx)
[1-f_N(\pp)-f_N(|\p-\q|)]. \label{gammasmallsinglet}
\end{eqnarray}
Inserting (\ref{gammasmallsinglet}) into (\ref{momdepNprod}), we obtain
\begin{eqnarray}
\partial_t f_{N\p}&=&2\frac{\tilde{\Gamma}_0}{m_\phi}\frac{M_\phi^2}{\p^2}
\int_{|M_\phi^2/(4\pp)-\pp|}^{\infty} d\qq \frac{\qq}{\Omega_{\phi\q}}
\left[1 -f_N(\pp) - f_N(\Omega_{\phi\q}-\pp)\right] 
\left[f_{\phi\q}-\bar f_{\phi\q}\right]\nonumber\\
&=&2\frac{\tilde{\Gamma}_0}{m_\phi}\frac{M_\phi^2}{\p^2}
\int_{M_\phi^2/(4\pp)+\pp}^{\infty} d\Omega_{\phi\q}
\left[1 -f_N(\pp) - f_N(\Omega_{\phi\q}-\pp)\right] 
\left[f_{\phi\q}-\bar f_{\phi\q}\right].\label{sigletmomentutmdependentproduction}
\end{eqnarray}
These expressions are our main results for the case of singlet decay.
In the derivation of above results, 
we only made use of the relation (\ref{NegativefN}), which holds for any Majorana fermion. Therefore they 
hold for \emph{any} phase space distribution function of $N$-particles. In particular, the results do not rely on the validity of the ansatz (\ref{fNAnsatz}).

\begin{figure}
	\center
	\includegraphics[width=0.7\textwidth]{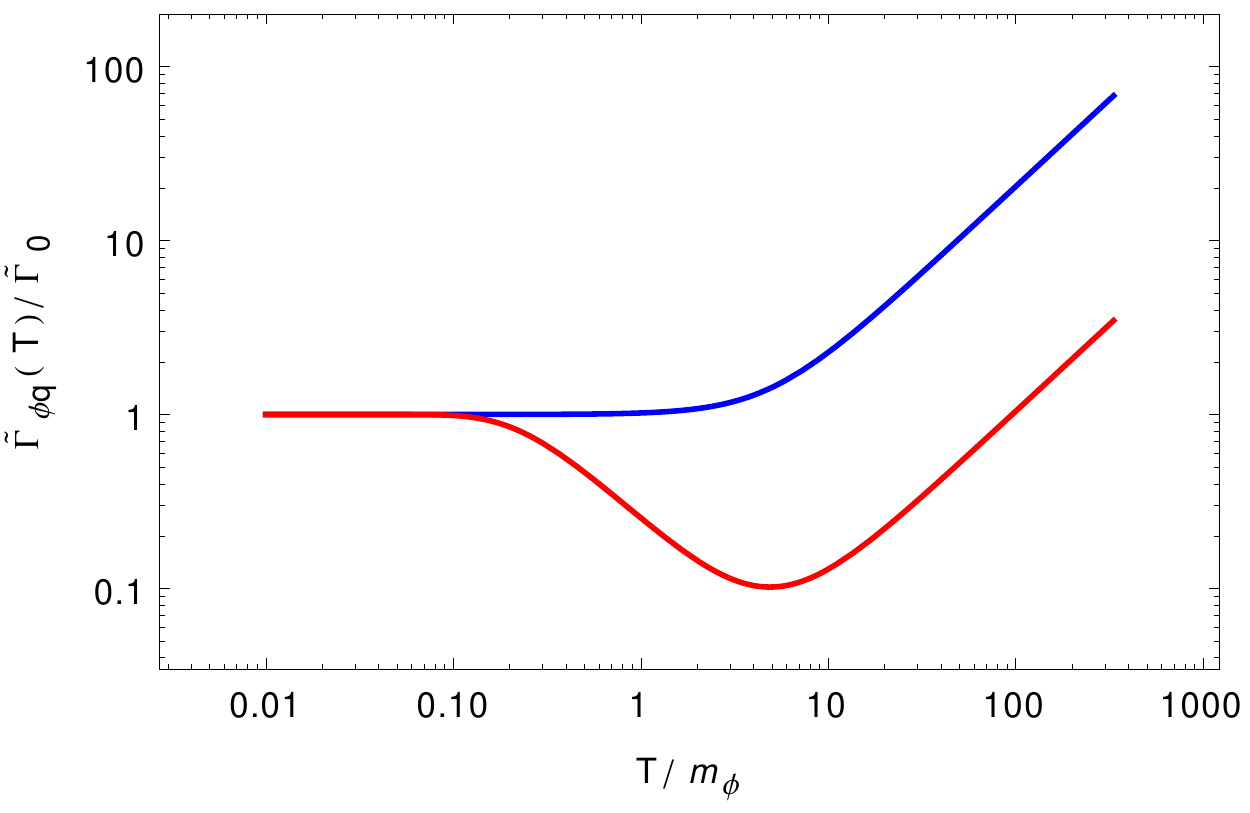}
\caption{The ratio $\tilde{\Gamma}_{\phi\q}/\tilde{\Gamma}_0$ for $\qq=0$, with $M_\phi$ given by (\ref{ThermalMass}) and $\lambda=1$, $\newbeta=1$, as a function of $T$. The blue curve is for $\newalpha=0$, the red curve for $\newalpha=1$.
The enhancement of the rate due to the increasing thermal $\phi$-mass kicks in when $M_\phi/m_\phi\gg 1$ above $T\simeq m_\phi\sqrt{24/\lambda}\simeq 5m_\phi$. For $\newalpha=1$, it competes with a Pauli suppression, which reduces the rate for $T>m_\phi$. 
\label{SingletZeroModeLamda1}}
\end{figure}

\begin{figure}
	\center
	\includegraphics[width=0.7\textwidth]{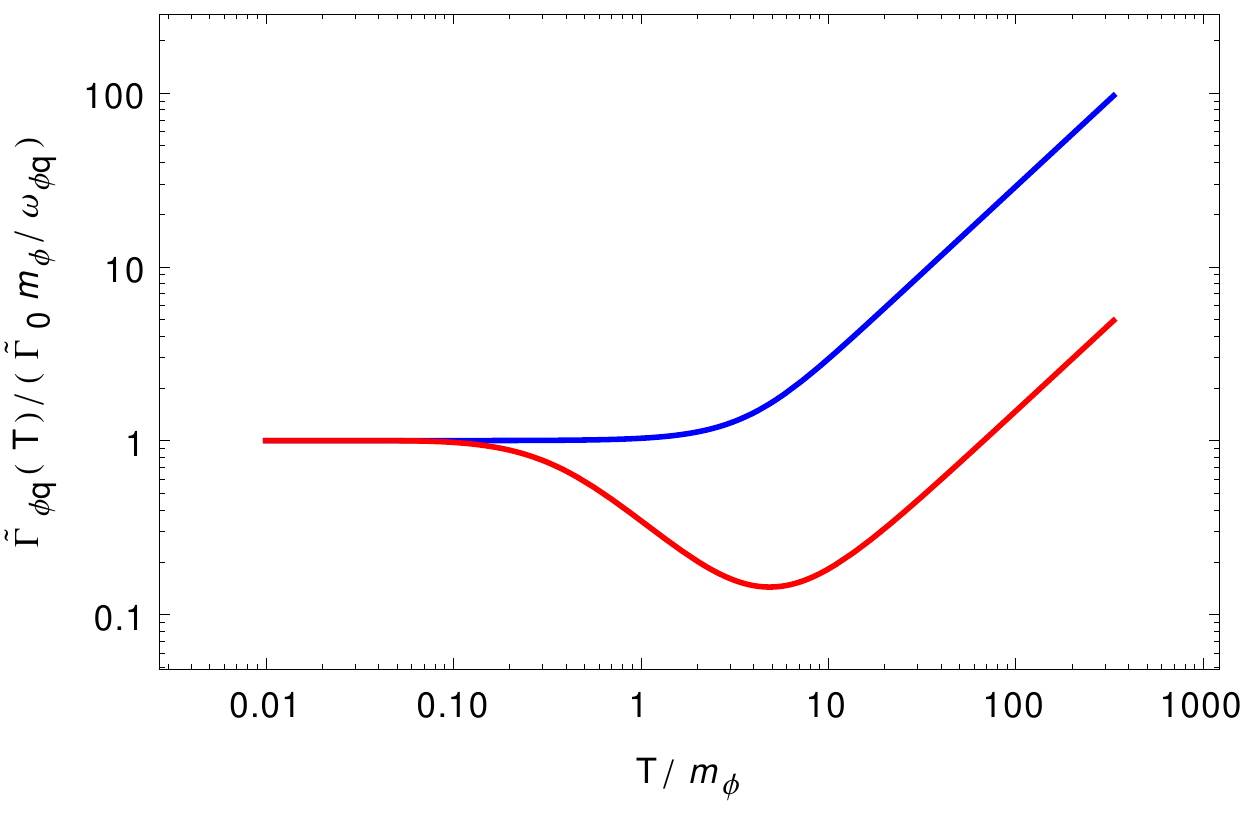}
\caption{The rate $\tilde{\Gamma}_{\phi\q}$ in units of the time-dilated rate $\tilde{\Gamma}_0 m_\phi/\omega_{\phi\q}$ for $\qq=m_\phi$, with $M_\phi$ given by (\ref{ThermalMass}) and $\lambda=1$, as a function of $T$. The blue curve is for $\newalpha=0$, the red curve for $\newalpha=1$.
The enhancement of the rate due to the increasing thermal $\phi$-mass kicks in when $M_\phi/m_\phi\gg1$ above $T\simeq m_\phi\sqrt{24/\lambda}\simeq 5m_\phi$. For $\newalpha=1$, it competes with a Pauli suppression, which reduces the rate for $T>m_\phi$. 
\label{SingletModeLamda1}}
\end{figure}

If the ansatz (\ref{fNAnsatz}) provides a valid approximation to the phase space distribution of $N$-particles, then the final integral can be solved analytically, 
\begin{eqnarray}\label{SingletDampingResultat}
\tilde{\Gamma}_{\phi\q}&=&\frac{y^2}{16\pi}
\frac{M_\phi^2}{\Omega_{\phi \q}}
\left[
1-2\newalpha\left(
1
-\frac{T}{\newbeta \qq}{\rm log}\left[
\frac{
f_F(\newbeta (\Omega_{\phi \q}-\qq)/2)
}{
f_F(\newbeta (\Omega_{\phi \q}+\qq)/2)
}
\right]
\right)
\right]\nonumber\\
&=&\tilde{\Gamma}_0\frac{M_\phi}{m_\phi}
\frac{M_\phi}{\Omega_{\phi \q}}
\left[
1-2\newalpha\left(
1
-\frac{T}{\newbeta \qq}{\rm log}\left[
\frac{
f_F(\newbeta (\Omega_{\phi \q}-\qq)/2)
}{
f_F(\newbeta (\Omega_{\phi \q} + \qq)/2)
}
\right]
\right)
\right]
.\end{eqnarray}
For $\newalpha=\newbeta=1$, this can be compared to the results found in \cite{Yokoyama:2004pf,Drewes:2010pf,Drewes:2013iaa,Ho:2015jva}.
Our result shows that, as long as the $N$-occupation numbers are far below their equilibrium value ($\newalpha\ll 1$), thermal effects actually enhance the production rate at high temperature because $M_\phi\gg m_\phi$ for $ T\gg m_\phi/\sqrt{\lambda}$.
Once some amount of DM has been produced, $\newalpha=0$ does not exactly hold. 
One may wonder if the statistical factors $f_N$ can have an effect towards the end of the DM production. Moreover, it is interesting to see what effect a pre-existing $N$-abundance may have. 
The considerations following (\ref{DMfraction}) show that the effect of $f_N$ is always negligible for a standard thermal history and if the average $N$-momentum is similar to $T$ ($\newbeta\simeq 1$).
For $\newbeta\gg 1$, there is room for a sizable $\newalpha$, and Pauli blocking may be significant for some modes. 
This either requires a production mechanism that lead to a rather ``cold'' pre-existing $N$-population or a significant injection of entropy into the primordial plasma after $N$-decoupling.
Some results are plotted in Figs.~\ref{SingletZeroModeLamda1}-\ref{SingletqT}.
\begin{figure}
	\center
	\includegraphics[width=0.7\textwidth]{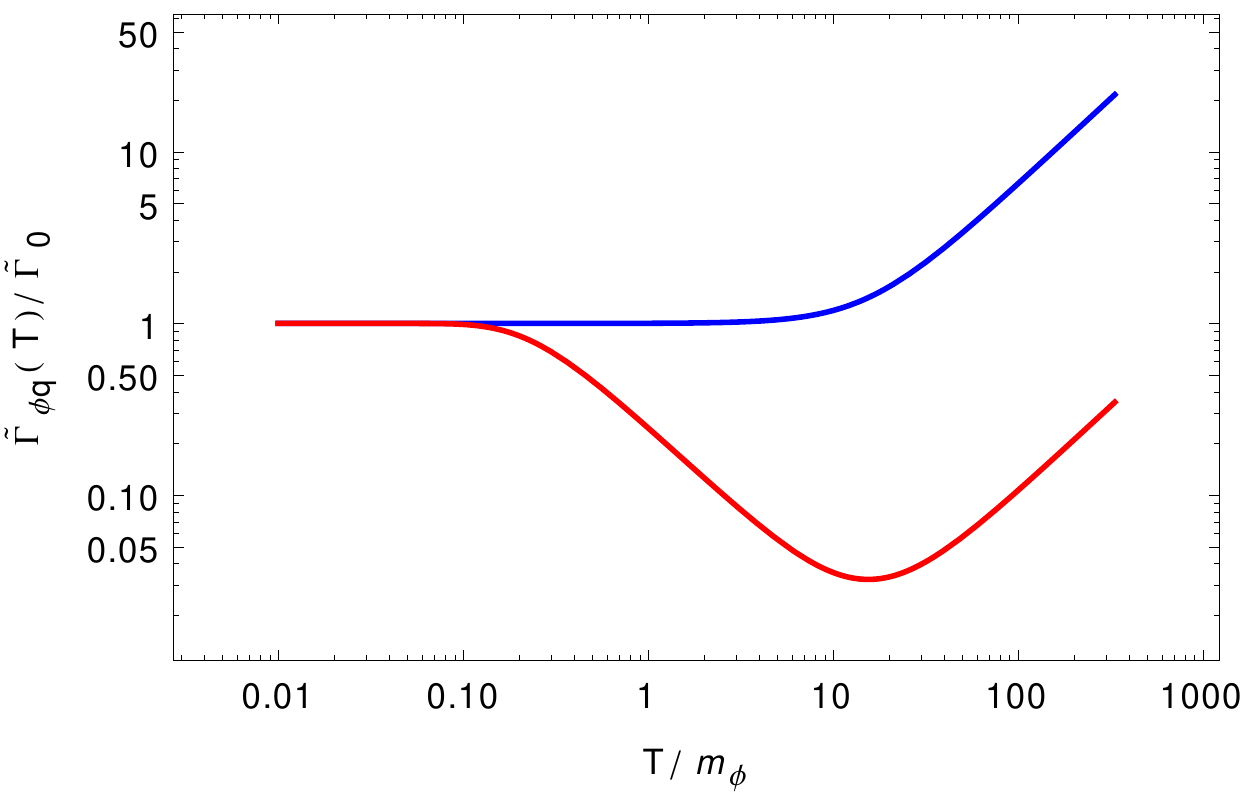}
\caption{The ratio $\tilde{\Gamma}_{\phi\q}/\tilde{\Gamma}_0$ for $\qq=0$, with $M_\phi$ given by (\ref{ThermalMass}) and $\lambda=0.1$ and $\newbeta=1$, as a function of $T$. The blue curve is for $\newalpha=0$, the red curve for $\newalpha=1$.
Comparison to Fig.~\ref{SingletZeroModeLamda1} shows the effect of changing $\lambda$.
\label{SingletZeroModeLamda01}}
\end{figure}
\begin{figure}
	\center
	\includegraphics[width=0.7\textwidth]{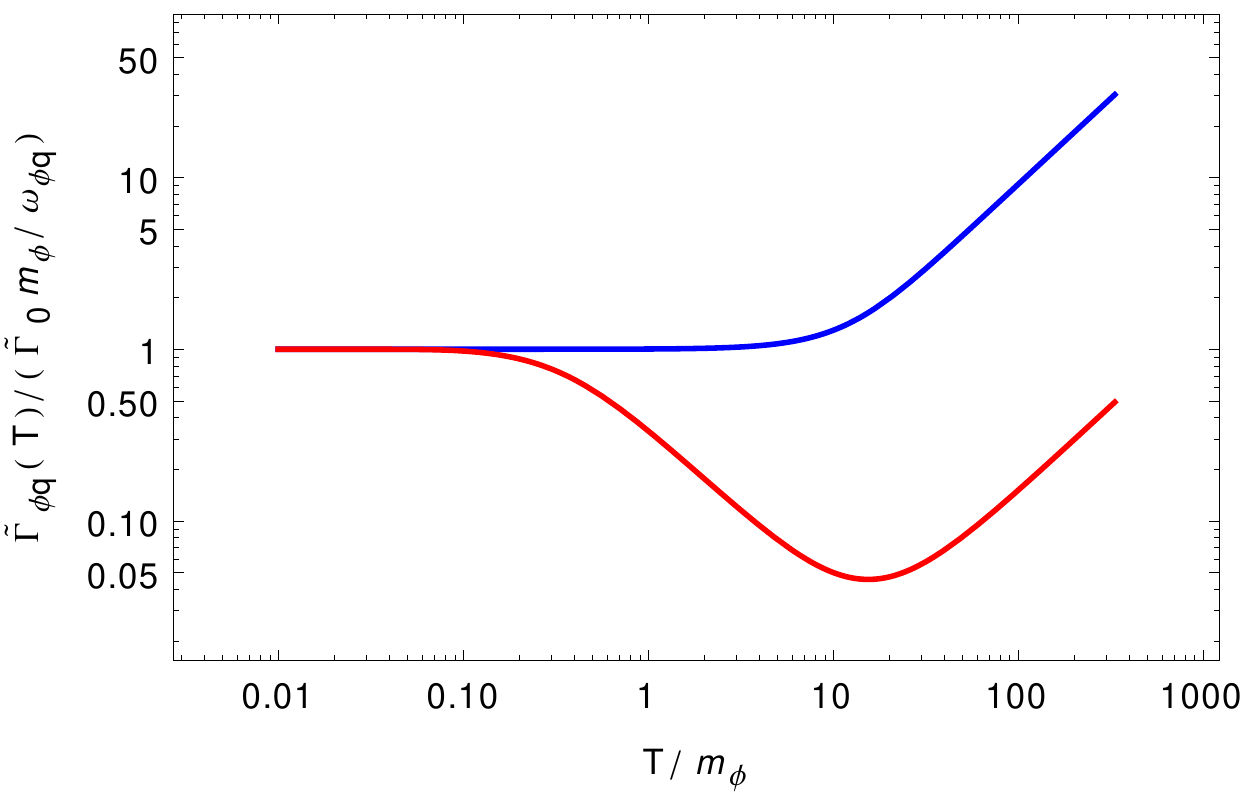}
\caption{The rate $\tilde{\Gamma}_{\phi\q}$ in units of the time-dilated rate $\tilde{\Gamma}_0 m_\phi/\omega_{\phi\q}$ for $\qq=m_\phi$, $M_\phi$ given by (\ref{ThermalMass}) and $\lambda=0.1$ as a function of $T$. The blue curve is for $\newalpha=0$, the red curve for $\newalpha=1$.
Comparison with Fig.~\ref{SingletModeLamda1} shows the effect of changing $\lambda$.
\label{SingletZeroModeLamda01}}
\end{figure}

In addition to (\ref{singletproduction}) there are other contributions to $\tilde{\Gamma}_{\phi\q}$ at order $y^2$, but these always contain additional powers of the coupling constants in $V(\phi)$ and $\mathcal{L}_{\phi {\rm int}}$ as well as loop factors. For instance, the potential (\ref{ScalarPotential}) gives rise to diagrams of the form
\begin{center}
\includegraphics[width=0.3\textwidth]{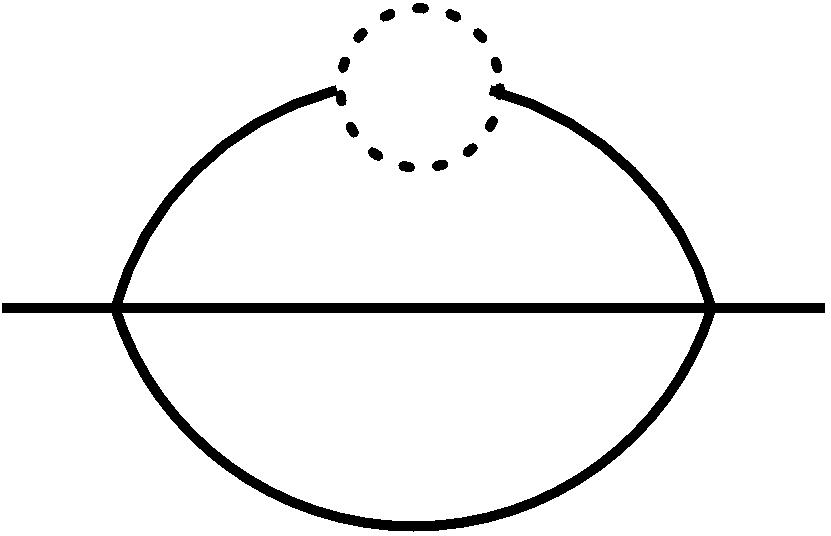},
\end{center}
cuts through which e.g. include processes $\phi\phi\rightarrow\phi N N$.
The size of these can be estimated by considering the equilibrium situation, in which case they are $y^2$-suppressed corrections to a ``setting sun'' contribution $\sim 10^{-3}\lambda^2T^2/\Omega_\q$ calculated in appendix B of \cite{Cheung:2015iqa}. Using (\ref{SingletDampingResultat}) and  (\ref{ThermalMass}), one can estimate that they are always suppressed by a factor $\ll\lambda$ relative to (\ref{SingletDampingResultat}) unless there is a very dense pre-existing population of $N$.

\subsection{Sizable Yukawa coupling $y$ and scatterings}  
The use of the free spectral density (\ref{rhoNfree}) is based on the assumption that the $N$-particles essentially do not feel the primordial plasma any more after they were produced. Since the $y\phi\bar{N}N$-vertex necessarily involves another $N$, this assumption is not only justified by the small Yukawa coupling ($y\ll1$), but also by the low density of $N$-particles ($f_N \ll 1$).
The use of (\ref{rhoNfree}) is only questionable if both of these suppressions are avoided, i.e., if $y$ is relatively large and there is a pre-existing $N$-population. We do not treat this special case here, but it is clear that one expects more significant thermal corrections to $\tilde{\Gamma}_{\phi\q}$. We expect these to be very similar to the thermal corrections found in section 7 of \cite{Drewes:2013iaa}. In that work, the produced fermion is a Dirac particle with gauge interactions. However, the thermal corrections to the spectral density of a Majorana fermion from Yukawa interactions \cite{Kiessig:2010pr} have the same structure as the gauge corrections for a Dirac fermion \cite{Weldon:1982bn}.
Finally, we would like to recall that (\ref{momdepNprod}) should only be used to determine $f_{N\p}$ if $N$-particles are predominantly produced in $1\rightarrow 2$ decays and do not re-scatter within a Hubble time. For sizable couplings $y$ this may not be the case. If scatterings are important, the momentum dependent $N$-production rate should directly be computed from an appropriate nonequilibrium generalisation of (\ref{FermionGamma}).

\section{Charged scalar decay}\label{ChargedSec}
Let us now consider the decay of a charged scalar $\Phi$.
To be specific, we use a model described by the Lagrangian\footnote{If $A_\mu$ is identified with the SM-gauge field, the $\frac{1}{4}F_{\mu\nu}F^{\mu\nu}$-term should of course be absorbed into $\mathcal{L}_{SM}$.}
\begin{eqnarray}
	\label{Lgauge}
	\mathcal{L} &=&\mathcal{L}_{SM}+ 
	\frac{1}{2} \overline{N}(i\slashed{\partial}-m_M)N-
	\overline{\ell_{L}}F N\tilde{h} -
	\tilde{h}^{\dagger}\overline{N}F^{\dagger}\ell_L 
+i\bar{\Psi}\Slash{\partial}\Psi
-\upalpha\bar{\Psi}\gamma^\mu A_\mu\Psi -\frac{1}{4}F_{\mu\nu}F^{\mu\nu}
\nonumber\\
&&+(\partial_\mu-i\upalpha A_\mu)\Phi^\dagger(\partial^\mu+i\upalpha A^\mu)\Phi - V(\Phi^\dagger\Phi) - \tilde{y}\left(\Phi\bar{N}\Psi + \bar{\Psi}N\Phi^\dagger\right)+\mathcal{L}_{\Phi {\rm int}}.
		\end{eqnarray}
Dark Matter is produced by the Yukawa interaction 
\begin{eqnarray}\label{Lcharged}
\tilde{y}\Phi\bar{\Psi}N + h.c.
\end{eqnarray}
The case where $\Phi$ is part of an additional Higgs doublet $h_\nu$ and carries SU(2) charge \cite{Adulpravitchai:2015mna} can be treated in the same way. In this case $\Psi$ should be identified with the lepton doublet $\ell_L$, and there are several massive scalars that decay.
The decay of each of them can be treated equivalently to the present calculation. 
The only differences are that $\ell_L$ is a chiral field (leading to a factor $1/2$ in the production rate) and the fact that the leptons in the final state carry SU(2) charge. The way how SU(2) interactions modify the spectral density $\uprho(p)$ is the same as for U(1), except for a numerical factor in the thermal fermion mass $\fermionmass{M}_f$ defined below \cite{Klimov:1981ka,Weldon:1982bn}.

\subsection{Heavy neutrino production}
The  kinetic equations for the occupation numbers 
during charged scalar decays can be obtained from \eqref{KineticEQ}-\eqref{KEQ} 
by the replacements $\phi \to \Phi$, $y \to \tilde{y}$ and $\Upomega_{N \p-\q} \to \Upomega_{\Psi \p-\q}$, where $\Upomega_{\Psi \p-\q}$ is the energy of a $\Psi$-quasiparticle with spatial momentum $\p-\q$.
The overall factor 2 in front of the integral in the kinetic equations \eqref{RateEquation}, \eqref{momdepNprod} and \eqref{KEQ} should be dropped because only one $N$-particle is produced in decays $\Phi\rightarrow N\Psi$.
This yields
\begin{eqnarray}\label{RateEquationPhi}
\partial_t n_N  = \int\frac{d^3\q}{(2\pi)^3} \tilde{\Gamma}_{\Phi\q} \left[f_{\Phi\q} -\bar f_{\Phi\q}\right],
\end{eqnarray}
\begin{eqnarray}\label{momdepNprodPhi}
\partial_t f_{N\p} = \int\frac{d^3\q}{(2\pi)^3}
\tilde{\gamma}_{\Phi}(\p,\q)\left[f_{\Phi\q}-\bar f_{\Phi\q} \right]\delta(\Omega_{\Phi \q} - \Upomega_{N \p} - \Upomega_{\Psi \p-\q}),  
\end{eqnarray}
where $\tilde{\gamma}_{\Phi}(\p,\q)$ is given from
\begin{eqnarray}\label{DefOfgammaPhi}
\tilde{\Gamma}_{\Phi\q}=\int\frac{d^3\p}{(2\pi)^3}\tilde{\gamma}_{\Phi}(\p,\q) \delta(\Omega_{\Phi \q} - \Upomega_{N \p} - \Upomega_{\Psi \p-\q}).
\end{eqnarray}
The rates $\tilde{\Gamma}_{\Phi \q}$ and $\tilde{\Gamma}_{\Phi\q}^\gtrless$ for $\Phi$ are defined analogously to the singlet case.
As in the case of the singlet scalar we will approximate $\Upomega_{N \p} \simeq \omega_{N \p} = (\p^2 + m_N^2)^{1/2}$ assuming that the coupling of $N$ to the plasma is very feeble. 

The fact that there is a charged particle in the final state makes the decay $\Phi\rightarrow N \Psi$ different from $\phi\rightarrow NN$ considered in the previous section.
First, the $\Psi$-quasiparticles in the final state tend to be thermally populated in the early universe (with a Fermi-Dirac distribution $f_F$ for the occupation numbers) due to their gauge interactions, and one expects significant Pauli blocking even if the $N$ population is low ($f_N \ll1$). 
Second, a charged particle feels the presence of the primordial plasma. This leads to ``screening", and the properties of $\Psi$-quasiparticles in the plasma are not the same as those of particles in vacuum. 
Finally, the gauge interactions of $\Psi$ open up new production channels for $N$. For example, there are s-channel and t-channel scatterings $\gamma \Phi \rightarrow N \Psi$ with intermediate $\Phi$ or $\Psi$ and ``photons'' in the initial state as well as their inverse processes.
When $N$-particles are produced in decays $\Phi\rightarrow N\Psi$, their momentum distribution can still be calculated using Boltzmann equation (\ref{momdepNprodPhi}). If scatterings play a significant role, (\ref{momdepNprodPhi}) cannot be used, and it is more convenient to directly calculate the $N$-self-energies $\Sigma_N^\gtrless(p)$ to obtain $f_{N\p}$. However, our present approach can still be used to calculate the total DM density from a rate equation (\ref{RateEquationPhi}).

The thermal corrections can be incorporated systematically by calculating $\tilde{\Pi}_\Phi^-(q)$ to a given order in perturbation theory. 
In the following we calculate $\tilde{\Pi}_\Phi^-(q)$ only to leading order in the small coupling $\tilde{y}$, but we use resummed \emph{hard thermal loop} (HTL) spectral densities for $\Psi$. This is necessary because the ``naive" loop expansion is not a consistent expansion in $\upalpha$ at high temperatures \cite{Linde:1980ts,Braaten:1989mz}.
While the use of resummed propagators for the interacting $\Psi$ is mandatory, we can (assuming that $f_N$ and $\tilde{y}$ are not too large) continue to use the free spectral density for the singlet $\uprho_N(p)$ given by (\ref{rhoNfree}).
The results obtained in this way are accurate to leading order in $\tilde{y}$ and leading $\log$ in the gauge coupling $\upalpha$.

\paragraph{$\Psi$ spectral density} - In the hard thermal loop (HTL) approximation the fermion spectral density reads \cite{Klimov:1981ka,Weldon:1982bn}
\begin{equation}
\uprho_\Psi(p)=\frac{1}{2}\left((\gamma_{0}-\hat{\bf{p}}\pmb{\gamma})\uprho_{+}+(\gamma_{0}+\hat{\bf{p}}\pmb{\gamma})\uprho_{-}\right)\label{HTLrho}
.\end{equation}
Here $\hat{\bf{p}}\pmb{\gamma}=p_{i}\gamma_{i}/|\textbf{p}|$. The two functions 
\begin{equation}
\uprho_{\pm}(p)\simeq2\pi[\uprho_{\pm}^{\rm pole}(p)+\uprho^{\rm cont}_{\pm}(p)]
\end{equation}
are the sum of singular contributions $\uprho_{\pm}^{\rm pole}$ and a continuous part $\uprho^{\rm cont}_{\pm}$.
The poles given by the singular parts define the energies (or dispersion relations) $\Upomega_{\pm}$ of quasiparticles with momentum $\p$,  
\begin{equation}\label{polepart}
\uprho_{\pm}^{\rm pole}(p)= Z_{\pm}\delta(p_{0}-\Upomega_{\pm})+Z_{\mp}\delta(p_{0}+\Upomega_{\mp}).
\end{equation}
The continuous part is given by
\begin{eqnarray}\label{continuouspart}
\lefteqn{\uprho^{\rm cont}_{\pm}(p)=\theta(1-x^{2})\frac{y^{2}}{|\textbf{p}|}(1\mp x)}\nonumber\\
&\times&\Bigg[\Bigg(1\mp x\pm y^{2}\left((1\mp x)\ln\left|\frac{x+1}{x-1}\right|\pm 2\right)\Bigg)^{2}
+\pi^{2}y^{4}(1\mp x)^{2}\Bigg]^{-1}.
\end{eqnarray}
Here $x=p_{0}/|\textbf{p}|$ and $y=\frac{1}{2}\fermionmass{M}_{f}/|\textbf{p}|$. 
The thermal fermion mass reads \cite{Weldon:1982bn}
\begin{equation}
\fermionmass{M}_{f}=\upalpha \frac{\rm C}{2} T.\label{fermionmassdef}
\end{equation} 
Here ${\rm C}$ is the quadratic Casimir of the representation of the gauge group.
For an U(1) interaction as considered here ${\rm C}=1$.
If $\Psi$ carries some kind of hypercharge $Y\neq 1$, one has to replace $\upalpha\rightarrow Y\upalpha$.
For SU(N) interactions the shape of $\uprho_\Psi$ is be exactly the same as for U(1), except for the numerical factor ${\rm C}$ in the definition of $\fermionmass{M}_f$ that depends on the representation under which $\Psi$ transforms.  
The mass $\fermionmass{M}_{f}$ is sometimes referred to as the {\it asymptotic mass}, it differs  from the plasma frequency $\upomega_f$ at $|\textbf{p}|=0$ by a factor $\sqrt{2}$.
The residues $Z_\pm$ are
\begin{equation}
Z_{\pm}=\frac{\Upomega_{\pm}^{2}-\textbf{p}^{2}}{
\fermionmass{M}_f^2
}
.\end{equation}
The dispersion relations $\Upomega_{+}$ and $-\Upomega_{-}$ have to be found as the solutions to
\begin{equation}
0=p_{0}-|\textbf{p}|\left[1+y^2\left((1-x)\ln\frac{x+1}{x-1}+2\right)\right]
.\end{equation}
There are two solutions, corresponding to two types of quasiparticles, with dispersion relations $\Upomega_+$ and $\Upomega_-$.  The former can be interpreted as a screened one-particle state, the latter are collective excitations \cite{Klimov:1982bv}. They are often referred to as ``holes'' or ``plasminos''.
There is an analytic expression for the dispersion relations $\Upomega_\pm$ in terms of the Lambert $W$-function \cite{Kiessig:2010pr}
\begin{equation}
\Upomega_+=|\textbf{p}|\frac{W_{-1}(s)-1}{W_{-1}(s)+1} \ , \ \Upomega_-=-|\textbf{p}|\frac{W_{0}(s)-1}{W_{0}(s)+1}
\end{equation}
with $s=-e^{-(y^{-2}+1)}$.

\paragraph{Self-energies} - 
The rate $\tilde{\Gamma}_{\Phi\q}$ at leading order in $\tilde{y}$ is given by the diagram 
\begin{center}
\includegraphics[width=0.3\textwidth]{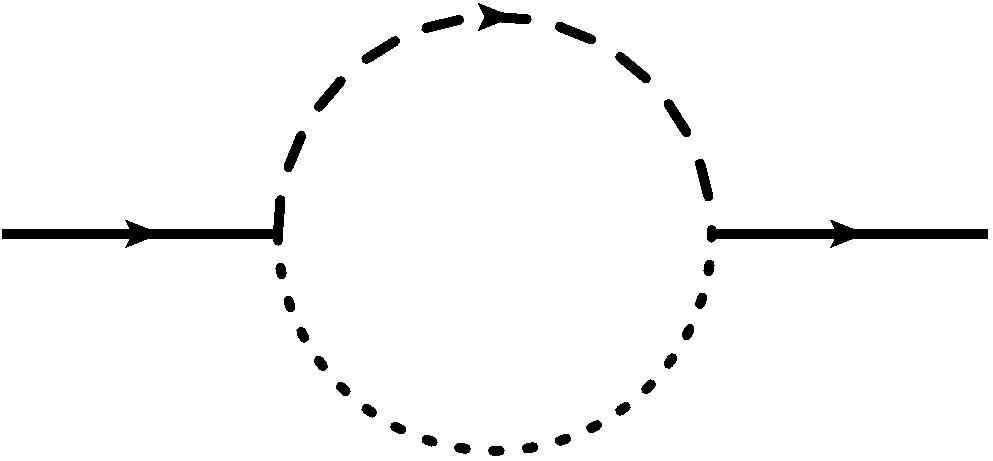},
\end{center}
where the solid lines represent the external $\Phi$, the dotted line is a free $N$-propagator and the dashed line a resummed $\Psi$-propagator. The arrow indicates charge flow.
The gain and loss terms read
\begin{eqnarray}
\tilde{\Pi}_\Phi^<(q)&=&i\tilde{y}^2\int\frac{d^4 p}{(2\pi)^4}{\rm tr}\left[S_N^<(p)S_\Psi^>(p-q)\right]\nonumber\\
&=&-i\tilde{y}^2\int\frac{d^4 p}{(2\pi)^4}\left(1-f_F(p_0-q_0)\right)f_N(p_0)
{\rm tr}\left[
\uprho_N(p)\uprho_\Psi(p-q)
\right] \label{doubletproduction<}
\end{eqnarray}
and
\begin{eqnarray}
\tilde{\Pi}_\Phi^>(q)&=&i\tilde{y}^2\int\frac{d^4 p}{(2\pi)^4}{\rm tr}\left[S_N^>(p)S_\Psi^<(p-q)\right]\nonumber\\
&=&-i\tilde{y}^2\int\frac{d^4 p}{(2\pi)^4}\left(1-f_N(p_0)\right)f_F(p_0-q_0)
{\rm tr}\left[
\uprho_N(p)\uprho_\Psi(p-q)
\right].
\end{eqnarray}
They combine into 
\begin{eqnarray}\label{doubletproduction}
\tilde{\Pi}_\Phi^-(q)=i\tilde{y}^2\int\frac{d^4 p}{(2\pi)^4}\left(f_N(p_0)-f_F(p_0-q_0)\right){\rm tr}\left[
\uprho_N(p)\uprho_\Psi(p-q)
\right].
\end{eqnarray}
The evaluation of \eqref{doubletproduction<}-(\ref{doubletproduction}) is complicated by the fact that the angle between the spatial vectors $\p$ and $\q$ appears in the complicated function $\uprho_\Psi(p-q)$. This can be avoided by making a shift in the spatial part $\p$ of the integration variable $p$ by $\q$ and introducing $\k=\p-\q$.
Afterwards we perform the integration over the angular variable $\xx=\q\k/(\qq\kk)$ between $\q$ and $\k$ with the help of $\delta(p^2-m_N^2)$ in (\ref{rhoNfree}).
The requirement that the zero of the $\delta$-function must be in the interval $\xx \in [-1,1]$ fixes the limits of the $p_0$ integral to 
\begin{equation}
\omega_\pm=\sqrt{(\qq\pm \kk)^2+m_N^2},
\end{equation}
where we have again used the notation $\qq=|\textbf{q}|$ and $\kk=|\textbf{k}|$.
The other angular integration is trivial because of rotational invariance.
Then it is straightforward to obtain
\begin{eqnarray}
\tilde{\Pi}_\Phi^>(q)&=&\frac{- i\tilde{y}^2}{2(2\pi)^2\qq}\int_0^\infty d\kk\int_{\omega_-}^{\omega_+}dp_0 \Bigg[\bigg[
f_F(p_0-q_0)\left[
1-f_N(p_0)\right] 
\nonumber\\
&&\phantom{X}\times\big[
\left(
\q^2+m_N^2-(p_0-\kk)^2
\right)
\uprho_{+}(p_0-q_0,\kk)\nonumber\\
&&\phantom{XX}+\left(
-\q^2-m_N^2+(p_0+\kk)^2
\right)
\uprho_{-}(p_0-q_0,\kk)
\big]\bigg]\nonumber\\
&&-\bigg[
p_0 \rightarrow -p_0
\bigg]
\Bigg],
\end{eqnarray}
\begin{eqnarray}
\tilde{\Pi}_\Phi^<(q)&=&\frac{-i\tilde{y}^2}{2(2\pi)^2\qq}\int_0^\infty d\kk\int_{\omega_-}^{\omega_+}dp_0 \Bigg[\bigg[
f_N(p_0) \left[1-f_F(p_0-q_0)
\right]\nonumber\\
&&\phantom{X}\times\big[
\left(
\q^2+m_N^2-(p_0-\kk)^2
\right)
\uprho_{+}(p_0-q_0,\kk)\nonumber\\
&&\phantom{XX}+\left(
-\q^2-m_N^2+(p_0+\kk)^2
\right)
\uprho_{-}(p_0-q_0,\kk)
\big]\bigg]\nonumber\\
&&-\bigg[
p_0 \rightarrow -p_0
\bigg]
\Bigg]
\end{eqnarray}
and therefore
\begin{eqnarray}\label{GeneralFormula}
\tilde{\Pi}_\Phi^-(q)&=&\frac{i\tilde{y}^2}{2(2\pi)^2\qq}\int_0^\infty d\kk\int_{\omega_-}^{\omega_+}dp_0 \Bigg[\bigg[\left[
f_N(p_0)-f_F(p_0-q_0)
\right]\nonumber\\
&&\phantom{X}\times\big[
\left(
\q^2+m_N^2-(p_0-\kk)^2
\right)
\uprho_{+}(p_0-q_0,\kk)\nonumber\\
&&\phantom{XX}+\left(
-\q^2-m_N^2+(p_0+\kk)^2
\right)
\uprho_{-}(p_0-q_0,\kk)
\big]\bigg]\nonumber\\
&&-\bigg[
p_0 \rightarrow -p_0
\bigg]
\Bigg].
\end{eqnarray}
For the $\qq=0$ mode, the $\xx$-integration in \eqref{doubletproduction<}-(\ref{doubletproduction}) is trivial, and the $\delta$-function can be used to perform the $p_0$-integral.
This yields
\begin{eqnarray}  \label{GeneralFormulaZeroMode21}
\tilde{\Pi}_\Phi^>(q)|_{\qq=0}&=&\frac{-i2\tilde{y}^2}{(2\pi)^2}
\int_0^\infty d\kk \frac{\kk^2}{\tilde{\omega}_\kk}
\bigg[ f_F(\tilde{\omega}_\kk-q_0)
\left[
 1-f_N(\tilde{\omega}_\kk)  
\right]\nonumber\\
&&\phantom{X}\times\left[
(\tilde{\omega}_\kk-\kk)\uprho_+(\tilde{\omega}_\kk-q_0,\kk)
+(\tilde{\omega}_\kk+\kk)\uprho_-(\tilde{\omega_\kk}-q_0,\kk)\right]\nonumber\\
&& -\Big[
q_0 \rightarrow - q_0 \Big] \bigg],
\end{eqnarray}
\begin{eqnarray}  \label{GeneralFormulaZeroMode12}
\tilde{\Pi}_\Phi^<(q)|_{\qq=0}&=&\frac{- i2\tilde{y}^2}{(2\pi)^2}
\int_0^\infty d\kk \frac{\kk^2}{\tilde{\omega}_\kk}
\bigg[
\left[
 1-f_F(\tilde{\omega}_\kk-q_0)\right]
f_N(\tilde{\omega}_\kk)  
\nonumber\\
&&\phantom{X}\times\left[
(\tilde{\omega}_\kk-\kk)\uprho_+(\tilde{\omega}_\kk-q_0,\kk)
+(\tilde{\omega}_\kk+\kk)\uprho_-(\tilde{\omega_\kk}-q_0,\kk)\right]\nonumber\\
&& -\Big[
q_0 \rightarrow - q_0 \Big] \bigg],
\end{eqnarray}
\begin{eqnarray}\label{GeneralFormulaZeroMode}
\tilde{\Pi}_\Phi^-(q)|_{\qq=0}&=&\frac{i2\tilde{y}^2}{(2\pi)^2}
\int_0^\infty d\kk \frac{\kk^2}{\tilde{\omega}_\kk}
\bigg[
\left[
f_N(\tilde{\omega}_\kk) - f_F(\tilde{\omega}_\kk-q_0)
\right]\nonumber\\
&&\phantom{X}\times\left[
(\tilde{\omega}_\kk-\kk)\uprho_+(\tilde{\omega}_\kk-q_0,\kk)
+(\tilde{\omega}_\kk+\kk)\uprho_-(\tilde{\omega_\kk}-q_0,\kk)\right]\nonumber\\
&&- \Big[
q_0 \rightarrow - q_0 \Big]
\bigg],
\end{eqnarray}
where $\tilde{\omega}_\kk= (m_N^2+\kk^2)^{1/2}$. Note that the above can be also obtained from more general expression \eqref{GeneralFormula} by taking limit $\qq \to 0$, which gives $\omega_+- \omega_- \approx 2 \qq \kk / \tilde{\omega}_\kk$ for the integral interval,  and by setting $p_0 =\tilde{\omega}_\kk$ in the integrand. In \eqref{GeneralFormulaZeroMode} the $\delta$-functions in $\uprho_{\pm}^{\rm pole}$ in principle allow to perform one more integration in the pole part analytically, but due to the appearance of $\Upomega_\pm$ it is in general not possible to find the zero analytically.

\subsection{Production in decays: Analytic approximations} 
For $M_\Phi\gg \fermionmass{M}_f+m_N$ the decay products have momenta $\sim M_\Phi/2$ that are large compared to the typical energy $\sim T$ of particles in the plasma. In this regime, one can approximate
\begin{eqnarray}\label{hardKapprox}
Z_+\simeq 1 \ , \ Z_-\simeq0 \ , \ \Omega_+^2\simeq \k^2 + \fermionmass{M}_f^2 \ , \ \uprho^{\rm cont}_{\pm}\simeq0 \ .
\end{eqnarray}
Physically this approximation of $Z_\pm$ means that the decay into holes can be neglected, which is intuitive because collective excitations are an infrared phenomenon that is only relevant for soft momenta.
The screened particles with hard momenta have a dispersion relation like free particles, but with the vacuum mass replaced by the momentum independent thermal mass $\fermionmass{M}_f$. This is also intuitive because the dispersion relation must be simple if the momentum exceeds all other scales in the problem, hence one expects that it is given by a constant mass term due to forward scattering.
Neglecting the continuous part of $\uprho_\Psi$ means that we only consider the contribution to $\tilde{\Gamma}_{\Phi\q}$ from $1\rightarrow2$ decays and their inverse, which is reasonable at low $T$.

\paragraph{Self-energies} - 
With the approximations (\ref{hardKapprox}) the integrals in (\ref{GeneralFormulaZeroMode21})-(\ref{GeneralFormulaZeroMode}) can be solved analytically, i.e.
\begin{eqnarray} \label{doubletzeromodefull>}
\lefteqn{\tilde{\Pi}_\Phi^>(q)|_{\qq=0}\simeq\frac{-i\tilde{y}^2}{\pi}
\frac{Q q_0+\fermionmass{M}_f^2-m_N^2}{2 Q q_0+\fermionmass{M}_f^2-m_N^2}\,
\left[
1- f_N(Q) \right]\left[1- f_F(q_0-Q)
\right]}\nonumber\\
&&\times\left[Q^2-m_N^2+Q\sqrt{Q^2-m_N^2}
\right]\theta(q_0^2-(m_N+\fermionmass{M}_f)^2),
\end{eqnarray}
\begin{eqnarray}  \label{doubletzeromodefull<}
\lefteqn{\tilde{\Pi}_\Phi^<(q)|_{\qq=0}\simeq\frac{-i\tilde{y}^2}{\pi}
\frac{Q q_0+\fermionmass{M}_f^2-m_N^2}{2 Q q_0+\fermionmass{M}_f^2-m_N^2}\,
 f_N(Q) \,f_F(q_0-Q)
}\nonumber\\
&&\times\left[Q^2-m_N^2+Q\sqrt{Q^2-m_N^2}
\right]\theta(q_0^2-(m_N+\fermionmass{M}_f)^2),
\end{eqnarray}
\begin{eqnarray}\label{doubletzeromodefull}
\lefteqn{\tilde{\Pi}_\Phi^-(q)|_{\qq=0}\simeq\frac{-i\tilde{y}^2}{\pi}
\frac{Q q_0+\fermionmass{M}_f^2-m_N^2}{2 Q q_0+\fermionmass{M}_f^2-m_N^2}\left[
1-f_F(q_0-Q)- f_N(Q)
\right]}\nonumber\\
&&\times\left[Q^2-m_N^2+Q\sqrt{Q^2-m_N^2}
\right]\theta(q_0^2-(m_N+\fermionmass{M}_f)^2)
\end{eqnarray}
with
\begin{eqnarray}
Q&=&\frac{q_0^2+m_N^2-\fermionmass{M}_f^2}{2q_0}.
\end{eqnarray}
\eqref{doubletzeromodefull} shows that Pauli blocking becomes important as soon as $T$ exceeds the vacuum $\Phi$-mass $m_\Phi$ even if there are no $N$-particles ($\newalpha=0$). The reason is of course the Pauli blocking of the $\Psi$-particles in the final state.
If we set $q_0=\Omega_{\Phi\q}=M_\Phi$, we can also again observe the enhancement of the rate due to the thermal mass $M_\Phi$. 
Since $\Phi$ is charged, it does not only receive a mass correction from its self-interaction, but also from the gauge interaction, and we have to take
\begin{equation}\label{GaugeThermalMass}
M_\Phi^2=m_\Phi^2 + \left(\frac{\lambda}{24} + \frac{\upalpha^2}{4}\right)T^2. 
\end{equation}
The term $\upalpha^2T^2/4$ arises because $\Phi$ carries a U(1) charge. 
If we were dealing with SM leptons and a Higgs doublet, as considered in \cite{Adulpravitchai:2015mna}, the equivalent to $\fermionmass{M}_f$ would be
\begin{equation}
M_\ell^2=\frac{1}{16}\left(3g^2 + g'^2\right)T^2,
\end{equation}
and the Higgs would have the thermal mass
\begin{equation}
M_h^2=\frac{1}{16}\left(3g^2 + g'^2 + 8\lambda\right)T^2  \ + \ {\rm Yukawa} \ {\rm contributions},
\end{equation}
where the different constant factors arise from the hypercharges \cite{Cline:1993bd}, cf.\ discussion after (\ref{fermionmassdef}).
If $T$ is near or below the temperature where spontaneous symmetry breaking occurs, there are additional contributions that are proportional to the temperature dependent expectation value of the Higgs condensate $\propto\langle \tilde{h}^\dagger \tilde{h} \rangle = \langle h^\dagger h \rangle$ (and  $\propto\langle h_\nu^\dagger h_\nu\rangle$ in the leptophilic two Higgs doublet model).

The expressions (\ref{doubletzeromodefull>})-(\ref{doubletzeromodefull}) can be significantly simplified if one neglects $m_N$ and $\fermionmass{M}_f$, i.e. 
\begin{eqnarray}
\tilde{\Pi}_\Phi^>(q)|_{\qq=0}\simeq\frac{-i\tilde{y}^2}{4\pi}\,q_0^2
\left[
1-f_F(q_0/2)\right]\left[1- f_N( q_0/2)
\right],
\end{eqnarray}
\begin{eqnarray}
\tilde{\Pi}_\Phi^<(q)|_{\qq=0}\simeq\frac{-i\tilde{y}^2}{4\pi}\,q_0^2
\,f_F(q_0/2) \, f_N( q_0/2)
\end{eqnarray}
and
\begin{eqnarray}
\tilde{\Pi}_\Phi^-(q)|_{\qq=0}\simeq\frac{-i\tilde{y}^2}{4\pi}\,q_0^2
\left[
1-f_F(q_0/2)- f_N( q_0/2)
\right].
\end{eqnarray}
Note that the assumption of massless final states is only justified if $M_\Phi\gg\fermionmass{M}_f+m_N$ at all relevant temperatures, which in the large $T$ limit is guaranteed by (\ref{GaugeThermalMass}) and (\ref{fermionmassdef}) only if $\lambda$ is sizable and $\Psi$ does not have additional interactions. 
\begin{figure}
	\center
	\includegraphics[width=0.7\textwidth]{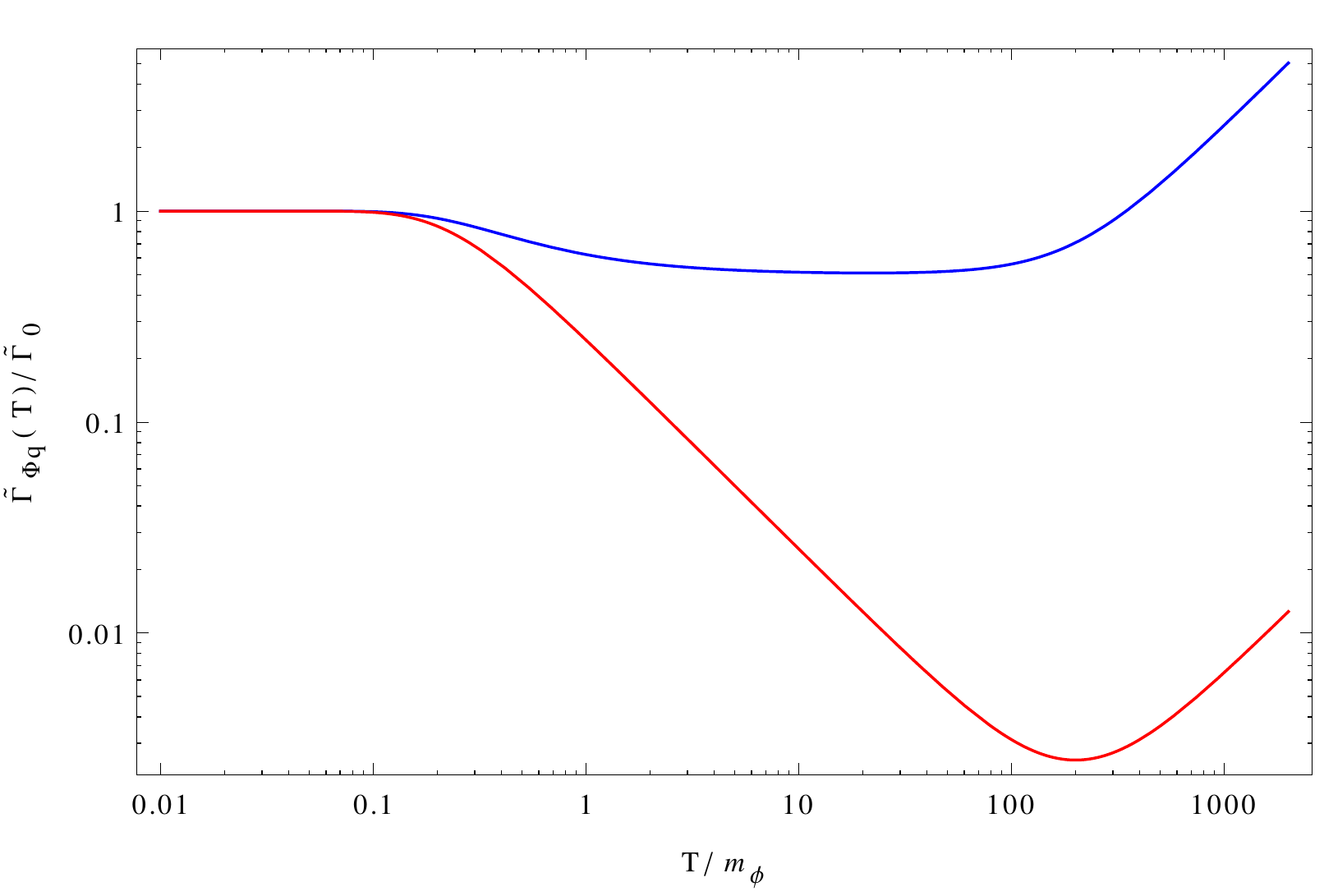}\\
\includegraphics[width=0.7\textwidth]{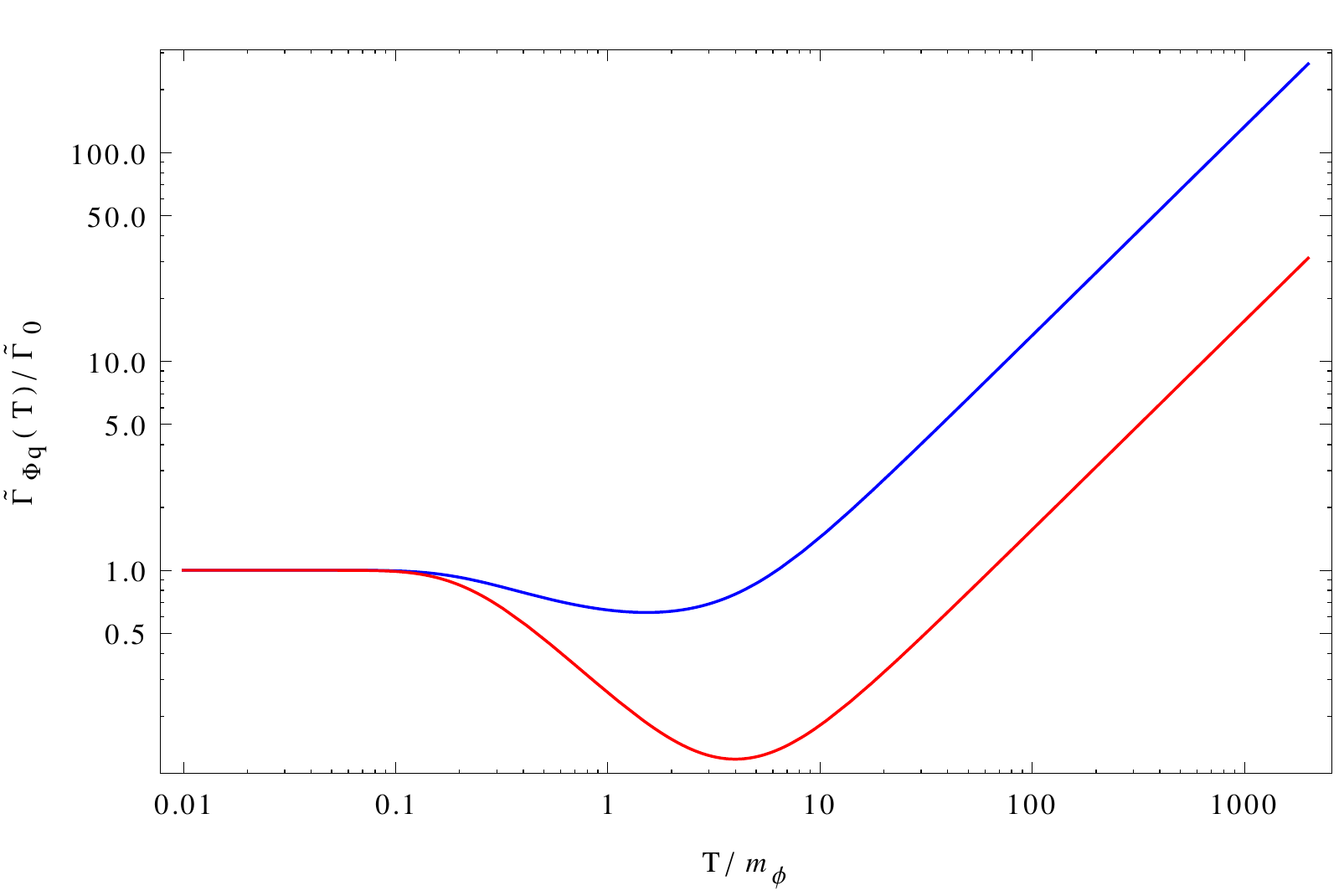}
\caption{The ratio $\tilde{\Gamma}_{\Phi\q}/\tilde{\Gamma}_0$ in the approximation (\ref{ZeroModeResultPsi}) for $\qq=0$, with $\newbeta=1$ and $M_\Phi$ given by (\ref{GaugeThermalMass}) and $\lambda=0$,  as a function of $T$. The blue curve is for $\newalpha=0$, the red curve for $\newalpha=1$.
In the upper plot the gauge coupling is chosen as $\upalpha=10^{-2}$, in the lower plot $\upalpha=1/2$. 
The enhancement of the rate due to the increasing thermal $\Phi$-mass is similar to the scalar case and would be even more prominent for $\lambda\neq0$. For $\newalpha=0$ the Pauli blocking can at most reduce $\tilde{\Gamma}_{\Phi\q}$ by a factor $1/2$, for $\newalpha>0$ a stronger suppression is possible.
\label{DoubletZeroMode}}
\end{figure}
As in the singlet case, it is not possible to find an analytic expression for $\tilde{\Gamma}_{\Phi\q}$ with $\q\neq0$ without additional assumptions about the phase space distribution function of $N$-particles. One can, however, again express $\tilde{\Gamma}_{\Phi\q}$ in terms of a one-dimensional integral.
In the approximation $m_N=\fermionmass{M}_f=0$ of massless final states and (\ref{hardKapprox}), the only contributing kinematic $\delta$-function is the one corresponding to the decay $\Phi\rightarrow N \Psi$ and we find
\begin{eqnarray}
\tilde{\Pi}_\Phi^>(q)=\frac{-i\tilde{y}^2}{4\pi\qq}\int_{(q_0-\qq)/2}^{(q_0+\qq)/2}d\kk 
\left[1- f_N(q_0-\kk)\right] \left[1-f_F(\kk) \right]\,
(q_0^2-\q^2),
\end{eqnarray}
\begin{eqnarray}
\tilde{\Pi}_\Phi^<(q)=\frac{-i\tilde{y}^2}{4\pi\qq}\int_{(q_0-\qq)/2}^{(q_0+\qq)/2}d\kk 
\,f_N(q_0-\kk) \,f_F(\kk)\, 
(q_0^2-\q^2),
\end{eqnarray}
\begin{eqnarray}\label{PiExpressionForDoublet}
\tilde{\Pi}_\Phi^-(q)=\frac{-i\tilde{y}^2}{4\pi\qq}\int_{(q_0-\qq)/2}^{(q_0+\qq)/2}d\kk 
\left[1-f_F(\kk)- f_N(q_0-\kk)
\right]
(q_0^2-\q^2),
\end{eqnarray}
where the integration limits have again been fixed by the usual kinematic considerations.

\paragraph{Production rate} - 
From these results, we finally obtain
\begin{eqnarray} \label{ZeroModeResultPsi21}
\tilde{\Gamma}_{\Phi\q}^>|_{\qq=0}\simeq
\tilde{\Gamma}_0
\frac{M_\Phi}{m_\Phi}
\left[
1-f_F(M_\Phi/2)\right]\left[1- f_N( M_\Phi/2)
\right],
\end{eqnarray}
\begin{eqnarray} 
\tilde{\Gamma}_{\Phi\q}^<|_{\qq=0}\simeq
\tilde{\Gamma}_0
\frac{M_\Phi}{m_\Phi}
\,f_F(M_\Phi/2) \, f_N( M_\Phi/2),
\end{eqnarray}
\begin{eqnarray}\label{ZeroModeResultPsi}
\tilde{\Gamma}_{\Phi\q}|_{\qq=0}\simeq
\tilde{\Gamma}_0
\frac{M_\Phi}{m_\Phi}
\left[
1-f_F(M_\Phi/2)- f_N( M_\Phi/2)
\right].
\end{eqnarray}
As in the previous section, we have expressed the thermal damping rate in terms of the vacuum damping rate $\tilde{\Gamma}_0$. 
Note that, compared to the singlet case, there is a symmetry-factor 2 difference in the latter, i.e., in this section we use
\begin{equation}
\tilde{\Gamma}_0 = \frac{\tilde{y}^2 m_\Phi}{8 \pi}
\end{equation}
instead of (\ref{Gammaphi0Def}).
For $\q\neq0$ we find
\begin{eqnarray}\label{GammaqExpressionForDoublet21}
\tilde{\Gamma}_{\Phi\q}^>=\frac{\tilde{y}^2}{8\pi}\frac{\Omega_{\Phi\q}^2-\q^2}{\Omega_{\Phi\q} \qq}\int_{(\Omega_{\Phi\q}-\qq)/2}^{(\Omega_{\Phi\q}+\qq)/2}d\kk 
\left[1- f_N(\Omega_{\Phi\q}-\kk)
\right]\left[1- f_F(\kk)\right],
\end{eqnarray}
\begin{eqnarray}\label{GammaqExpressionForDoublet12}
\tilde{\Gamma}_{\Phi\q}^<=\frac{\tilde{y}^2}{8\pi}\frac{\Omega_{\Phi\q}^2-\q^2}{\Omega_{\Phi\q} \qq}\int_{(\Omega_{\Phi\q}-\qq)/2}^{(\Omega_{\Phi\q}+\qq)/2}d\kk 
f_F(\kk)f_N(\Omega_{\Phi\q}-\kk),
\end{eqnarray}
\begin{eqnarray}\label{GammaqExpressionForDoublet}
\tilde{\Gamma}_{\Phi\q}=
\frac{\tilde{\Gamma}_0}{m_\Phi}
\frac{\Omega_{\Phi\q}^2-\q^2}{\Omega_{\Phi\q} \qq}\int_{(\Omega_{\Phi\q}-\qq)/2}^{(\Omega_{\Phi\q}+\qq)/2}d\kk 
\left[1-f_F(\kk)- f_N(\Omega_{\Phi\q}-\kk)
\right].
\end{eqnarray}
With the approximation $\Omega_{\Phi\q}^2\simeq M_\Phi^2+\q^2$, this reads 
\begin{eqnarray}\label{DoubletSimpleInterpretation}
\tilde{\Gamma}_{\Phi\q}=\tilde{\Gamma}_0\frac{M_\Phi}{m_\Phi}
\frac{M_\Phi}{\Omega_{\Phi\q}}\frac{1}{\qq}\int_{(\Omega_{\Phi\q}-\qq)/2}^{(\Omega_{\Phi\q}+\qq)/2}d\kk 
\left[1-f_F(\kk)- f_N(\Omega_{\Phi\q}-\kk)
\right].
\end{eqnarray}
The physical interpretation of this expression is very simple and the same as discussed following (\ref{singletGammaArbitraryfN2}).
From (\ref{DefOfgammaPhi}) we have
\begin{eqnarray}\label{tildeGammaPhi}
\tilde{\gamma}_{\Phi}(\p,\q)=\frac{\tilde{y}^2\pi}{q_0\omega_{N\p}}\left(\omega_{N \p}+\frac{\p^2-\p\q}{|\p-\q|}\right)
\left[1-f_N(\omega_{N\p}) -f_F(q_0-\omega_{N\p})\right].
\end{eqnarray}
This allows to derive an expression similar to (\ref{sigletmomentutmdependentproduction}) for the charged scalar case,
\begin{eqnarray}
\partial_t f_{N\p}&=&\frac{\tilde{y}^2}{4\pi}\int_{\Omega_1}^{\Omega_2}d\Omega_{\Phi\q} \frac{\Omega_{\Phi\q}-\omega_{N \p}}{\pp\,\omega_{N \p}}
\left[\omega_{N \p} +\frac{1}{2\tilde{\omega}}\left(\tilde{\omega}^2+\p^2-\q^2 \right) \right]\nonumber\\
&&\times\left[1 - f_N(\omega_{N \p}) - f_F(\Omega_{\Phi\q}-\omega_{N \p} )\right]
\left[f_{\Phi\q}-\bar f_{\Phi\q}\right]  \label{doubletmomentutmdependentproduction}
\end{eqnarray}
with $\tilde{\omega}\equiv\sqrt{(\Omega_{\Phi\q} - \omega_{N \p})^2-\fermionmass{M}_f^2}$  and  
\begin{eqnarray}
\Omega_2&=& \frac{\omega_{N \p}(M_\Phi^2+m_N^2 -\fermionmass{M}_f^2) + 
\pp \sqrt{\Big[(M_\Phi+\fermionmass{M}_f)^2 - m_N^2\Big] \Big[(M_\Phi-\fermionmass{M}_f)^2 - m_N^2\Big]}}{2 m_N^2} \, , \nonumber  \\ 
\Omega_1&=& {\rm Max}\left[M_\Phi ,   \frac{\omega_{N \p}(M_\Phi^2+m_N^2 -\fermionmass{M}_f^2) - 
\pp \sqrt{\Big[(M_\Phi+\fermionmass{M}_f)^2 - m_N^2\Big] \Big[(M_\Phi-\fermionmass{M}_f)^2 - m_N^2\Big]}}{2 m_N^2} \right] \, . \nonumber
\end{eqnarray}
In the limit $\fermionmass{M}_f=m_N=0$ of massless final states, (\ref{doubletmomentutmdependentproduction}) simplifies to
\begin{eqnarray}
\partial_t f_{N\p}&=&
\frac{\tilde{\Gamma}_0}{m_\phi}\frac{M_\Phi^2}{\p^2}
\int_{M_\phi^2/(4\pp)+\pp}^{\infty} d\Omega_{\Phi\q}
\left[1 -f_N(\pp) - f_F(\Omega_{\Phi\q}-\pp)\right] 
\left[f_{\Phi\q}-\bar f_{\Phi\q}\right],\label{doubletmomentutmdependentproductionsimple}
\end{eqnarray}
which can be compared to (\ref{sigletmomentutmdependentproduction}). 
Note that in contrast to the singlet case there is no factor 2 in front of the integral because only one $N$-particle is produced in each decay.

\begin{figure}
	\center
	\includegraphics[width=0.7\textwidth]{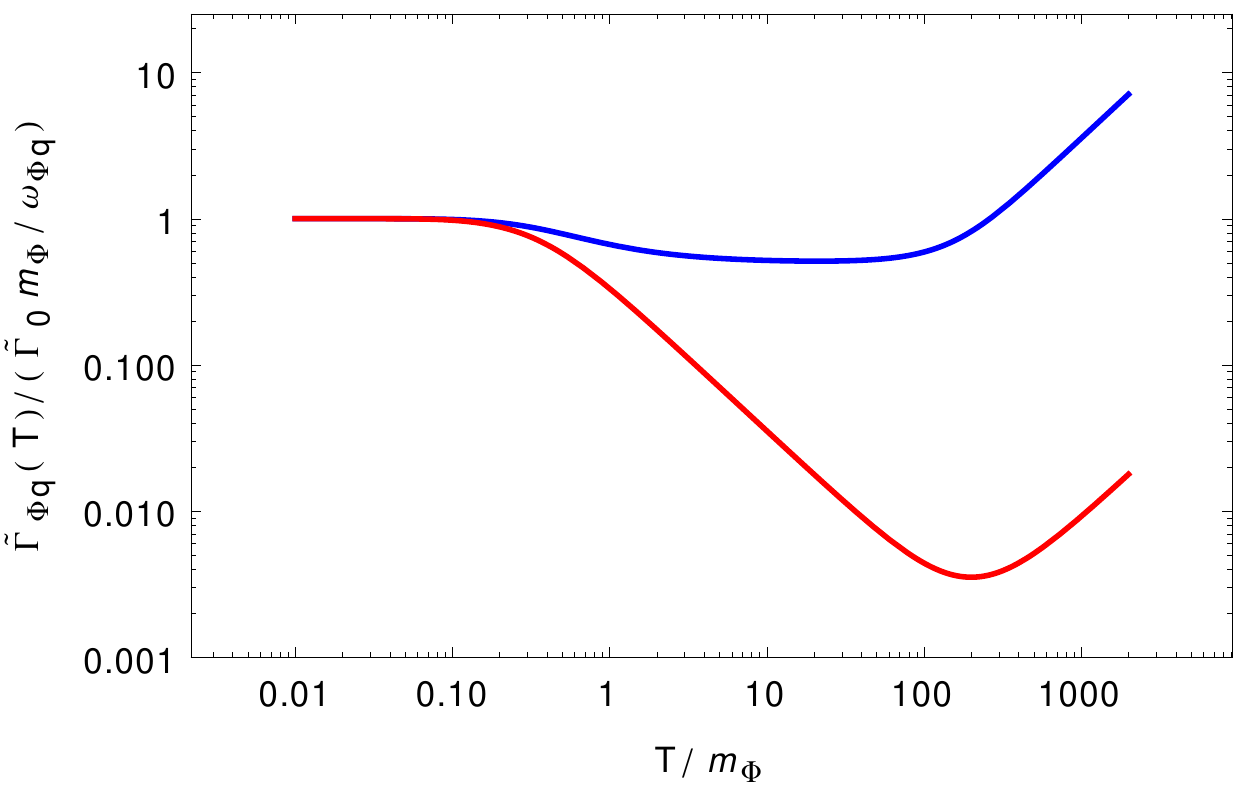}\\
\includegraphics[width=0.7\textwidth]{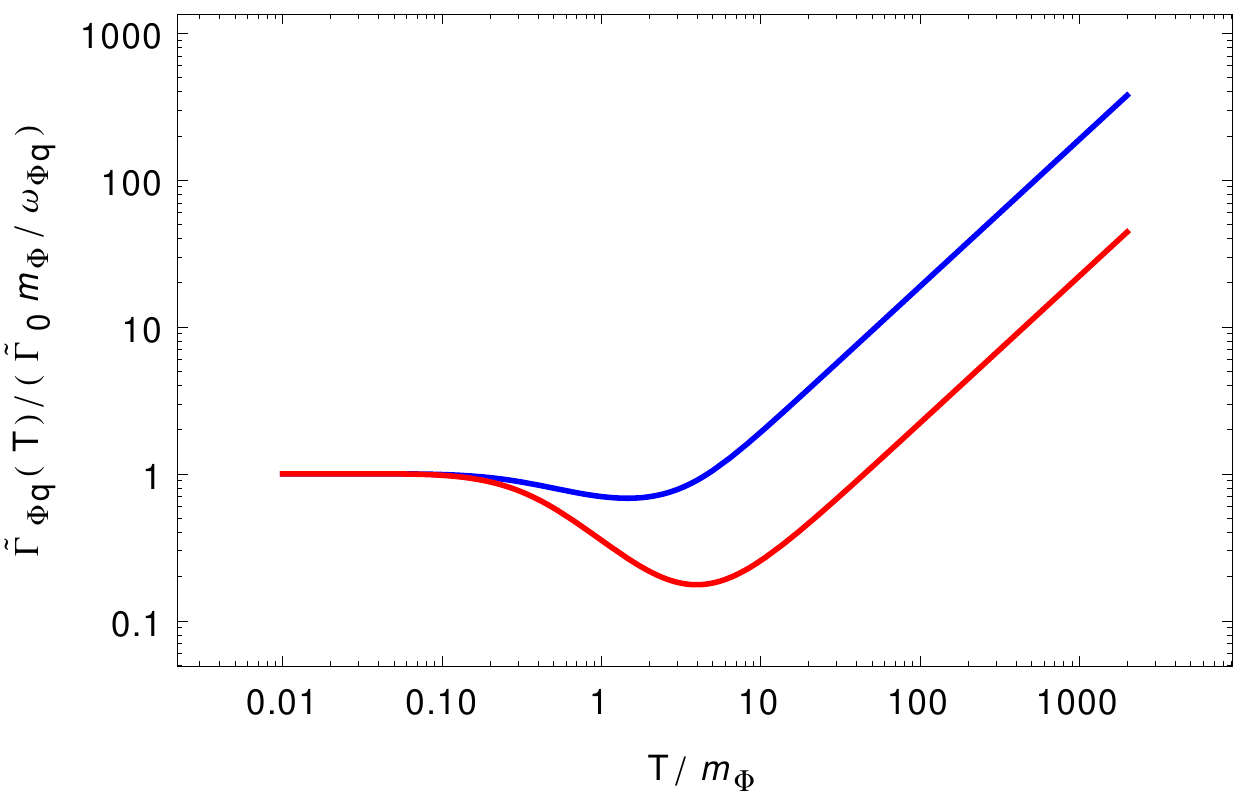}
\caption{The rate $\tilde{\Gamma}_{\Phi\q}$ in units of the time-dilated rate $\tilde{\Gamma}_0 m_\Phi/\omega_{\Phi\q}$ in the approximation (\ref{QnonzeroResult}) for $\qq=m_\Phi$, with $M_\Phi$ given by (\ref{GaugeThermalMass}) and $\lambda=0$, as a function of $T$. The blue curve is for $\newalpha=0$, the red curve for $\newalpha=1$.
In the upper plot the gauge coupling is chosen as $\upalpha=10^{-2}$, in the lower plot $\upalpha=1/2$. 
As in Fig.~\ref{DoubletZeroMode}, the enhancement of the rate due to the increasing thermal $\Phi$-mass is similar to the scalar case and would be even more prominent for $\lambda\neq0$. For $\newalpha=0$ the Pauli blocking can at most reduce $\tilde{\Gamma}_{\Phi\q}$ by a factor $1/2$, for $\newalpha>0$ a stronger suppression is possible.
\label{DoubletMode}}
\end{figure}

\begin{figure}
	\center
	\includegraphics[width=0.7\textwidth]{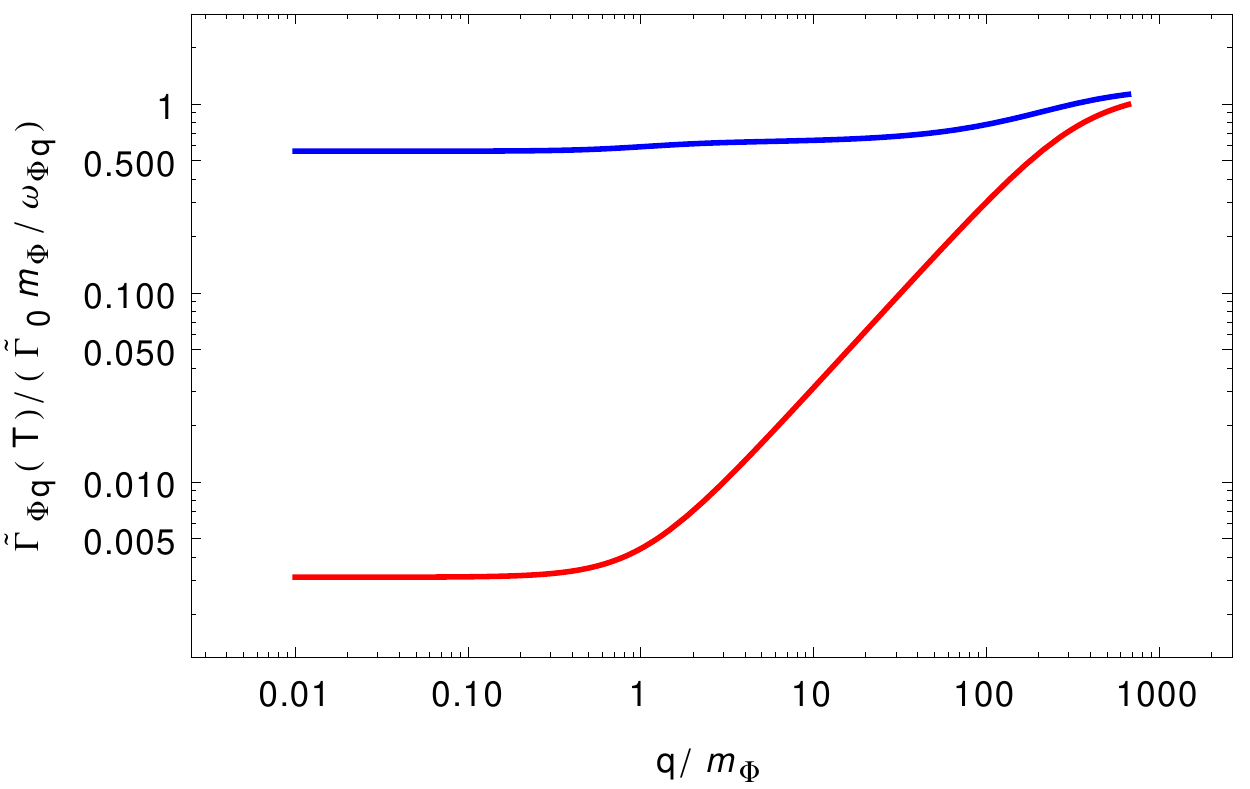}\\
\includegraphics[width=0.7\textwidth]{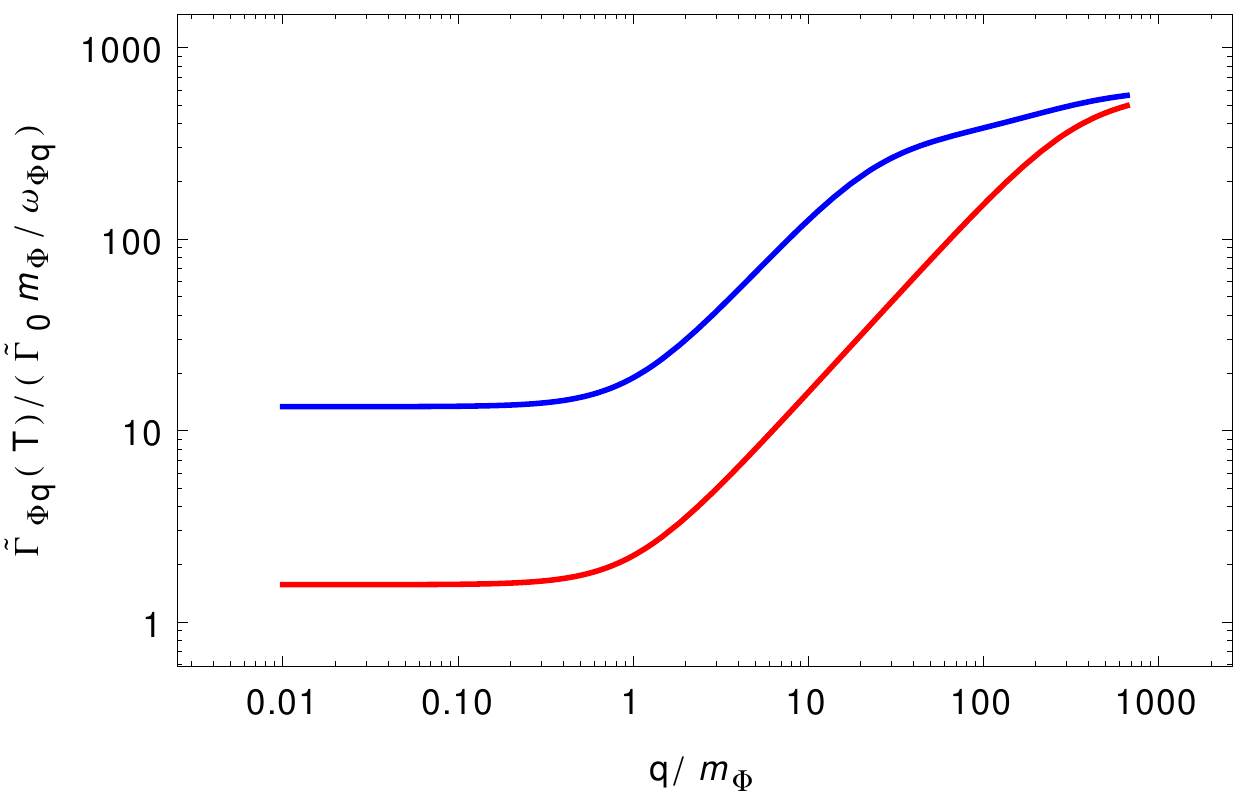}
\caption{The rate $\tilde{\Gamma}_{\Phi\q}$ in units of the time-dilated rate $\tilde{\Gamma}_0 m_\Phi/\omega_{\Phi\q}$ in the approximation (\ref{QnonzeroResult}) for $T=100m_\Phi$, with $M_\Phi$ given by (\ref{GaugeThermalMass}) and $\lambda=0$ as a function of $\qq$. The blue curve is for $\newalpha=0$, the red curve for $\newalpha=1$.
In the upper plot the gauge coupling is $\upalpha=10^{-2}$, in the lower plot $\upalpha=1/2$.
As usual, at high temperatures, the increased thermal mass $M_\Phi$ leads to a larger rate than the vacuum estimate $\tilde{\Gamma}_0 m_\Phi/\omega_{\Phi\q}$ suggests. The effect is larger for stronger gauge coupling and would be more prominent for $\lambda\neq0$.
\label{DoubletqT}}
\end{figure}

These expressions hold for an arbitrary sterile neutrino phase space distribution.
If one assumes that $f_N$ can be approximated by the ansatz (\ref{fNAnsatz}), then even the final integral  in \eqref{GammaqExpressionForDoublet} 
can be solved analytically for arbitrary $\q$, i.e.
\begin{eqnarray}\label{QnonzeroResult}
\tilde{\Gamma}_{\Phi\q}&\simeq&\tilde{\Gamma}_0
\frac{M_\Phi}{m_\Phi}
\frac{M_\Phi}{\Omega_{\Phi\q}}
\Bigg[
\frac{T}{\qq}\log\left[
\frac{f_F\left(
\frac{\Omega_{\Phi\q} - \qq}{2}
\right)
}
{f_F\left(
\frac{\Omega_{\Phi\q}+\qq}{2}
\right)
}
\right]
+\newalpha
\left[
-1 + 
\frac{T}{\newbeta\qq}\log\left[
\frac{f_F\left(\newbeta
\frac{(\Omega_{\Phi\q}-\qq)}{2}
\right)
}
{
f_F\left(\newbeta
\frac{(\Omega_{\Phi\q}+\qq)}{2}
\right)
}
\right]
\right]
\Bigg].
\end{eqnarray}
The different factors in the results (\ref{DoubletSimpleInterpretation}) and (\ref{QnonzeroResult}) have a simple physical interpretation, which is exactly analogue to the discussion following (\ref{singletGammaArbitraryfN2}). 
At low temperatures, they reproduce the $T=0$ decay rate. 
Once the temperature exceeds $m_\Phi$, Pauli-blocking of the $\Psi$-particle in the final state suppresses $\tilde{\Gamma}_{\Phi\q}$. This effectively reduces $\tilde{\Gamma}_{\Phi\q}$ (for $\q=0$, $T\gg M_\Phi$ and $f_N\ll1$ roughly by a factor $1/2$).
If there is a significant population of $N$-particles ($\newalpha \simeq 1$), then Pauli blocking also applies to the $N$ in the final state as soon as the temperature approaches $\sim\newbeta m_\Phi$.  
At even higher temperatures, when $M_\Phi/m_\Phi\gg1$, the thermal $\Phi$-mass increases the rate again due to the reduced lifetime of a heavier particle. For a potential of the form (\ref{ScalarPotential}) this roughly happens at  $T>m_\Phi(24/\lambda)^{1/2}$. 
This behaviour is illustrated in Figs.~\ref{DoubletZeroMode}-\ref{DoubletqT}.

Up to now we have completely neglected the contributions from the holes $\Upomega_-$  as well as the term $\uprho^{\rm cont}_{\pm}$. At $T\ll m_\Phi/\upalpha$ this is justified because $M_\Phi\gg\fermionmass{M}_f$, so the decay products' momenta $\pp\sim M_\Phi/2$ are hard with respect to the plasma. The holes $\Upomega_-$ are relevant only in a small temperature interval: The suppression of $Z_-$ is only lifted when the temperature is high enough that $\fermionmass{M}_f\sim M_\Phi$, but low enough to keep $M_\Phi>m_N + \upomega_f$ because otherwise the decay into holes is kinematically forbidden.    
Outside this small interval, we do not expect the holes $\Upomega_-$ to play an important role.

At this point one should compare the analytic expressions obtained in the singlet and charged case, i.e., (\ref{singletnonyeroresultArbitraryfN}) to (\ref{PiExpressionForDoublet}), (\ref{Gammasingletzeromode}) to (\ref{ZeroModeResultPsi}), (\ref{singletGammaArbitraryfN2}) to (\ref{GammaqExpressionForDoublet}) and (\ref{SingletDampingResultat}) to (\ref{QnonzeroResult}).
This comparison shows that, under the present assumptions, the production rates of sterile neutrinos in singlet and charged scalar decays look almost the same and are governed by the same physical effects, namely Pauli blocking and ``thermal masses'' in the plasma. 
The only differences are the occupation numbers (the charged particles in the final state of $\Phi$-decays are in thermal equilibrium) and the actual values of the thermal masses (which depend on the interactions that are responsible for forward scatterings and screening).
There is, however, another difference. In the singlet-model (\ref{Lsinglet}) with tiny $y$, $N$-particles are in very good approximation only produced in $\phi$-decays. The contribution to $\tilde{\Gamma}_{\phi\q}$ from scatterings is suppressed by additional powers of the coupling $y$ and $f_N$.\footnote{This of course assumes that the $N$ have no other interactions at the relevant temperatures. If the $N$ are charged under an additional gauge symmetry, this entirely changes the situation, and scatterings usually bring them into thermal equilibrium.}
Physically this means that $N$-particles effectively do not interact with the plasma any more after they have been produced.
Mathematically it implies that the free spectral density (\ref{rhoNfree}) is a good approximation for $\uprho_N$.
In the model (\ref{Lcharged}), on the other hand, the $\Phi$-particles and $\Psi$-particles are charged. At sufficiently large $T$, their number density is large enough that they scatter frequently, and one expects that $N$-particles can be produced in scatterings (rather than decays).
Mathematically this is reflected by the fact that the expression (\ref{HTLrho}) for $\uprho_\Psi$ does not only consist of the pole-contribution (\ref{polepart}), which is equivalent to (\ref{rhoNfree}), but also the continuous part (\ref{continuouspart}), which we have ignored so far.

\subsection{Full damping rate and scatterings}\label{Sec:VollesRohr}  

\begin{figure}
	\center
	\includegraphics[width=0.7\textwidth]{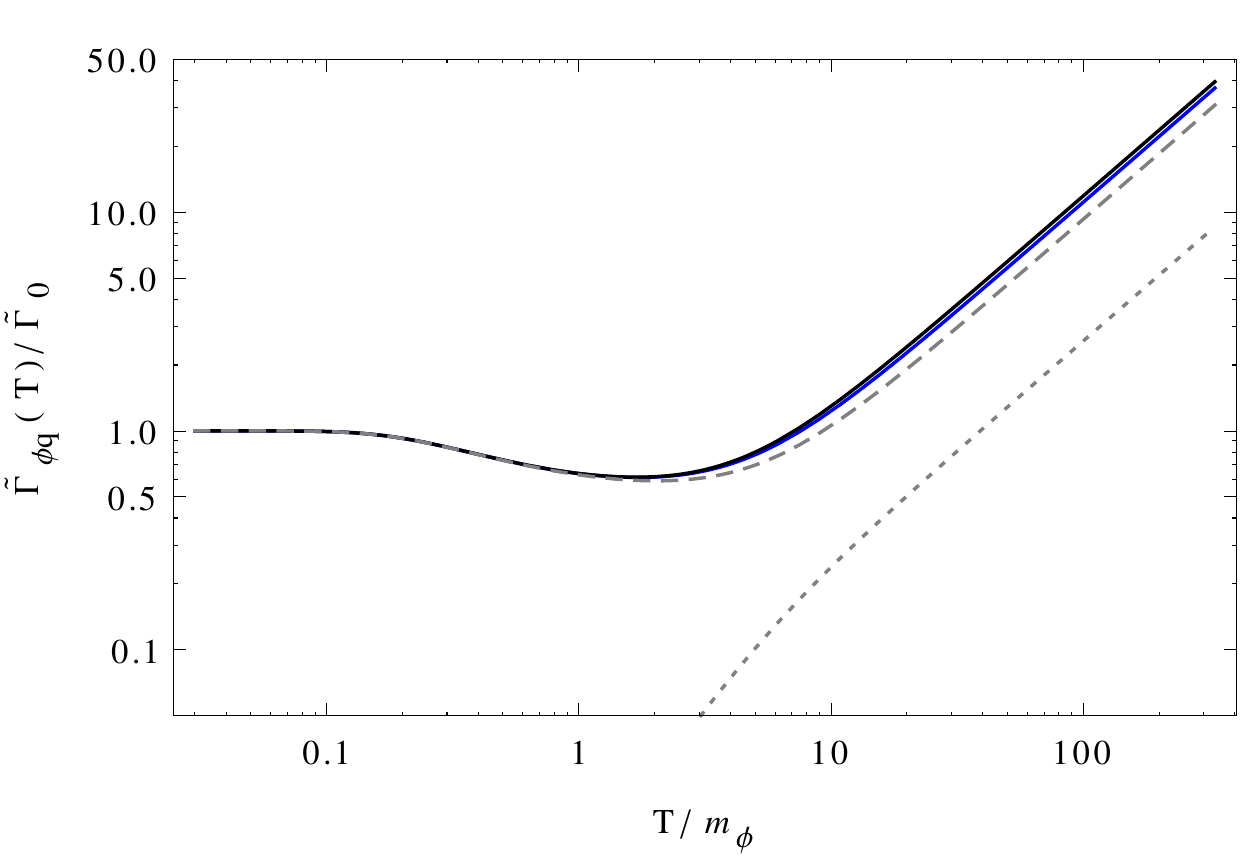}\\
\caption{The ratio $\tilde{\Gamma}_{\Phi\q}/\tilde{\Gamma}_0$ for $\qq=0$, with $M_\Phi$ given by (\ref{GaugeThermalMass}),  $\lambda=1$, $\upalpha=1/10$ and $\newalpha=0$ as a function of $T$. 
The solid black curve is obtained from numerical evaluation of (\ref{GeneralFormulaZeroMode}), the blue curve shows the approximation  (\ref{ZeroModeResultPsi}). 
The gray curves are the individual contributions to (\ref{GeneralFormulaZeroMode}) from the pole contributions $\uprho_\pm^{\rm pole}$ (dashed) and the continuum $\uprho_\pm^{\rm cont}$ (dotted) in $\uprho_\Psi$ as given in (\ref{HTLrho}).
\label{FullRateDoublet2}}
\end{figure}

\begin{figure}
	\center
	\includegraphics[width=0.7\textwidth]{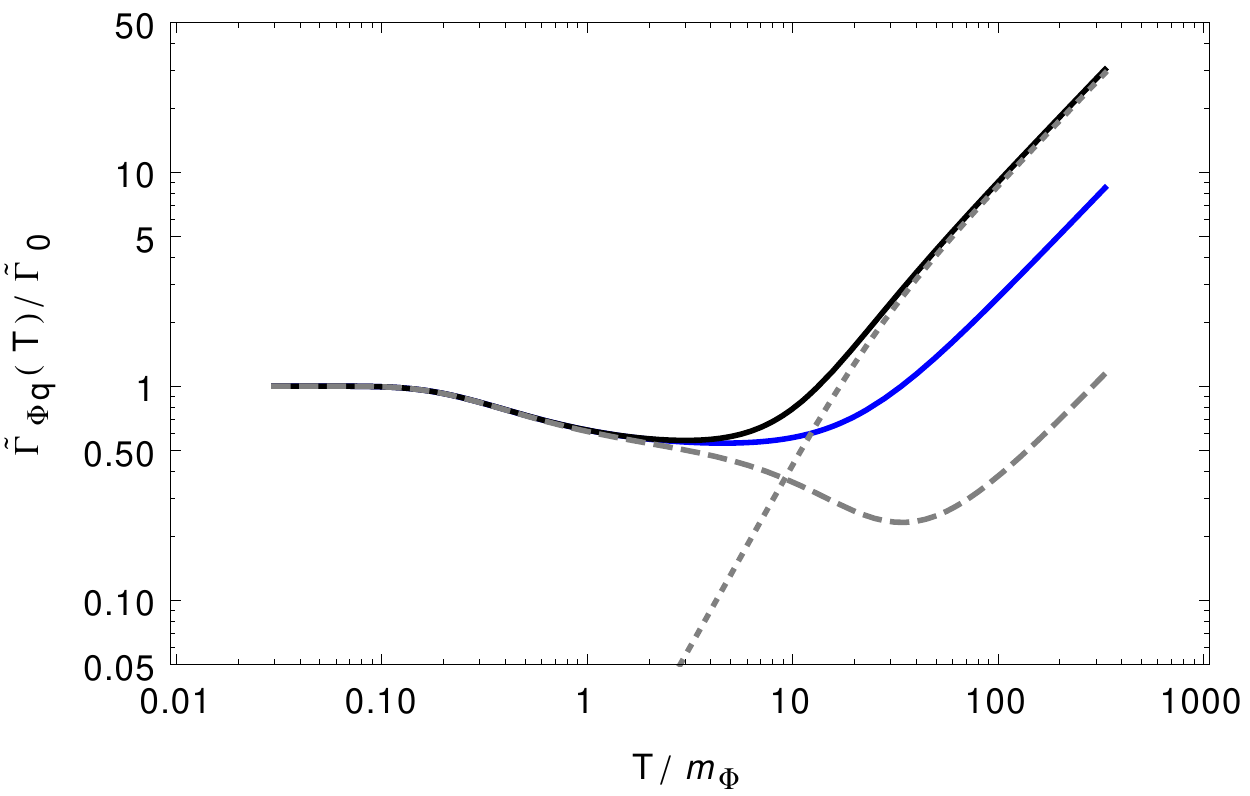}\\
\caption{Same as Fig.~\ref{FullRateDoublet2}, but for $\lambda=0$.
\label{FullRateDoublet}}
\end{figure}

As pointed out above, the decay $\Phi\rightarrow N \Psi$ is Pauli suppressed for $T>m_\Phi$, leading to a factor $1/2$ suppression of $\tilde{\Gamma}_{\Phi\q}$ for $f_N\ll1$ and a bigger suppression for $f_N\sim 1$.\footnote{For $M_\Phi < \fermionmass{M}_f$ the decay can in principle even become completely forbidden kinematically, but in our model this is not possible unless $m_N>m_\Phi$ because the thermal $\Phi$-mass corrections is always at least as big as the thermal $\Psi$-mass.}
On the other hand, the continuum contribution from $\uprho^{\rm cont}_{\pm}$ grows with temperature. It is strongly dominated by the $\uprho^{\rm cont}_{-}(\tilde{\omega}_\k-q_0)$-term.
This can be understood physically because this contribution comes from Landau damping due to scatterings, and the number of possible scattering partners in the bath increases. 
To incorporate all these effects, we have to solve the integral in (\ref{GeneralFormulaZeroMode}) numerically. The result is shown in Figs.~\ref{FullRateDoublet2} and \ref{FullRateDoublet}.

\paragraph{Regimes and scenarios} - One can clearly distinguish different temperature regimes. For $T\ll m_\Phi$, the vacuum decay rate is an excellent approximation. 
For $m_\Phi \lesssim T < m_\Phi/\upalpha$ the main effect of the plasma is the suppression of $\tilde{\Gamma}_{\Phi\q}$ due to Pauli blocking, and (\ref{ZeroModeResultPsi}) is a good approximation.
For 
$M_\Phi\gg m_\Phi$ the thermal correction (\ref{GaugeThermalMass}) to the $\Phi$-mass kicks in and enhances $\tilde{\Gamma}_{\Phi\q}$ again. (\ref{ZeroModeResultPsi}) remains to be a good approximation. What happens at larger temperatures depends on the value of $\lambda$.

Let us first set  $\lambda\gg \upalpha^2$. 
This situation is shown in  Fig.~\ref{FullRateDoublet2}.
In this case the hierarchy $M_\Phi\gg \fermionmass{M}_f$ holds at all temperatures because the thermal $\Phi$-mass is larger than the asymptotic $\Psi$-mass.
Then (\ref{hardKapprox}) can be applied even at large temperatures, (\ref{ZeroModeResultPsi}) remains a valid approximation at $T>m_\Phi/\upalpha$, and the discussion following (\ref{QnonzeroResult}) applies.
That is, (\ref{GammaqExpressionForDoublet}) and \eqref{doubletmomentutmdependentproduction}  can be used to compute $f_{N\p}$ for all temperatures if $\lambda\gg\upalpha^2$.\footnote{One may wonder why the decay approximation holds for arbitrarily high temperatures. Since the number density of scattering partners grows with $T$, one would expect that $\Phi$-annihilation in scatterings should dominate over $\Phi$-decay at sufficiently high temperature.  One physical explanation is that the $\Phi$-quasiparticles simply are too short lived due to their large thermal mass to find a scattering partner before they decay.
 For $\lambda=0$, the thermal mass and scattering rate are determined by the same gauge coupling constant $\upalpha$, and scatterings indeed become more efficient at some temperature, c.f. dotted line in figure \ref{FullRateDoublet}. For $\lambda\gg\upalpha$ the large thermal mass (\ref{GaugeThermalMass}) ensures that the decay rate always remains larger than the scattering rate.   
}

The situation $\lambda\ll \upalpha^2$ is shown in Fig.~\ref{FullRateDoublet}.
In this case, the ratio $\fermionmass{M}_f/M_\Phi$ approaches unity for large temperatures.
The approximations for $\Upomega_+$ and $Z_+$ in (\ref{hardKapprox}) start to break down at $T\sim m_\Phi/\upalpha$, which can be seen from the fact that the blue line in Fig.~\ref{FullRateDoublet} deviates from the gray dashed line. The reason is that (\ref{hardKapprox}) is based on the assumption that the produced $\Psi$-quasiparticles have large momenta $\pp\gg \fermionmass{M}_f=\upalpha T/2$ with respect to the plasma, which is only true for $M_\Phi\gg \fermionmass{M}_f, m_N$. 
If this is not fulfilled, the pole contribution becomes sensitive to the details of the fermion dispersion relations $\Upomega_\pm$ in the infrared. At the same time, the contribution from the continuum part $\uprho_\pm^{\rm cont}$, which is negligible for all lower temperatures, starts to dominate. 
This contribution can be interpreted as $\Phi$-annihilation in scatterings.
The resulting total rate $\tilde{\Gamma}_{\Phi\q}$ grows linear with $T$ (as expected due to the increasing number density of scattering partners), but with a coefficient that is larger than predicted by (\ref{ZeroModeResultPsi}). 
We have verified this for different choices of the parameters, but it is not clear at this stage if a simple analytic expression for the coefficient can be found. 
Note that Fig.~\ref{FullRateDoublet} shows the result for the zero-mode $\q=0$.  For $\q\neq0$ 
the validity of (\ref{hardKapprox}) extends to larger temperatures because the decay product's momenta in the plasma rest frame are larger.
For small values of $\lambda$, our simple results therefore hold only up to temperatures $T\sim m_\Phi/\upalpha$; at higher temperatures $\tilde{\Gamma}_{\Phi\q}$ is dominated by scatterings, and the integrals (\ref{GeneralFormulaZeroMode21})-(\ref{GeneralFormulaZeroMode}) have to be evaluated numerically. Since the $N$-momentum cannot be reconstructed from $\q$ in a scattering, one cannot use (\ref{doubletmomentutmdependentproduction}) to calculate $f_{N\p}$. One can, however, still use a rate equation (\ref{RateEquationPhi}) to compute the total DM density. 
If one is interested in the DM momentum distribution in the regime where $N$-particles are predominantly produced in scatterings, it would be more convenient to directly evaluate the fermionic self-energies $\Sigma_N^\gtrless$ with nonequilibrium propagators in the loop to determine the production rate from (\ref{FermionGamma}). This goes beyond the scope of the present work, which is focused on DM scalar decays.

\paragraph{A comment on scatterings} -
We have used resummed thermal $\Psi$-propagators to account for thermal corrections to $\tilde{\Gamma}_{\Phi\q}$.
This does not only allow to take quantum statistical factors and the correct quasiparticle dispersion relations in the decays $\Phi\rightarrow N\Psi$ into account, but also includes some of the contributions to $\tilde{\Gamma}_{\Phi\q}$ due to scatterings. 
These are encoded in the continuous part of $\uprho_\Psi$.
In the HTL-approximation (\ref{continuouspart}), this continuous part is non-zero only for $x^2<1$. 
The resulting contribution to $\tilde{\Gamma}_{\Phi\q}$ can be interpreted as Landau damping due to logarithmically enhanced t-channel scatterings \cite{Garbrecht:2013urw}. The full $\uprho_\Psi$ (beyond HTL) would also give a contribution for $x^2>1$, which corresponds to s-channel scatterings. These are, however, not logarithmically enhanced. Therefore the use of the HTL approximation corresponds to a systematic expansion in the gauge coupling to leading $\log$ in $\upalpha$ \cite{Garbrecht:2013urw}.

Additional contributions from scatterings have been studied by different authors \cite{Anisimov:2010gy,Besak:2012qm,Laine:2013lka,Biondini:2013xua,Garbrecht:2013gd,Garbrecht:2013urw,Ghisoiu:2014ena} for the case that $\Phi$ is identified with the SM-Higgs and $m_N\gg m_\Phi$ (while we assumed $m_N\ll m_\Phi$). 
In the high temperature regime $T\gg M_\Phi \gg m_N, m_\Phi$, the difference in the hierarchy of vacuum masses $m_N$ and $m_\Phi$ should be negligible because the thermal mass corrections dominate the kinematics.
A direct comparison is nevertheless difficult because the authors of the above works numerically calculated the inclusive rate $\Upgamma_N$ under the assumption that $\Phi$ is in equilibrium. This situation is e.g. realised in several leptogenesis scenarios. For the case of sterile neutrino DM production considered here, the phase space distribution $f_\Phi$ for $\Phi$ is a dynamic quantity that in general deviates from equilibrium (in particular in the ``freeze-in'' scenario). 
Even if we fix $f_\Phi$ to a Bose-Einstein distribution and obtain a total production rate from  (\ref{RateEquationPhi}), 
the result still depends on the choice of gauge group under which $\Phi$ is charged.
While it is straightforward to generalise the analytic result (\ref{QnonzeroResult}) we found from U(1) interactions to an arbitrary SU(N) gauge group by making the replacements discussed after (\ref{fermionmassdef}), there appears to be no simple way to obtain the contribution from multiple scatterings for general SU(N) interactions and non-equilibrium $\Phi$ from the numerical results in the literature.  
The discussion of the approximations made in \cite{Besak:2012qm,Garbrecht:2013urw}, however, suggests that the contributions from vertex- and ladder-diagrams are not logarithmically enhanced, and that our treatment is accurate to leading $\log$  order in the gauge coupling $\upalpha$.
In conclusion, the above considerations suggest that  calculation of $\tilde{\Gamma}_{\Phi\q}$ is accurate to order $\tilde{y}^2$ in the Yukawa coupling and order $\upalpha^2\log\upalpha^{-2}$ in the gauge coupling.

\section{Discussion and Conclusions}\label{Conclusions}
We have studied thermal corrections to the production rate of singlet fermions $N$ in the decay of neutral scalars $\phi$ and charged scalars $\Phi$ in a hot plasma. This rate determines the total abundance and momentum distribution of Dark Matter particles in scenarios where sterile neutrino Dark Matter is produced in the decay of heavier particles.
With some modifications that account for the different spin, our results may be generalised to the production of other DM candidates in decays, including gravitinos \cite{Moroi:1993mb,Giudice:1999am,Feng:2003xh,Cheung:2011nn,Roszkowski:2014lga} or WIMPs \cite{Moroi:1999zb,Kitano:2008tk,Acharya:2009zt}.

We found that the sterile neutrino production rate receives considerable thermal corrections if the plasma temperature $T$ exceeds the mass of the decaying scalar.  Our main results include expressions (\ref{sigletmomentutmdependentproduction}) and (\ref{doubletmomentutmdependentproduction}), which can be used to determine the abundance and momentum distribution of heavy neutrinos produced in decays. They are based on the gain- and loss rates (\ref{Gammasingletzeromode21})-(\ref{gammasmallsinglet}) for singlet decays and (\ref{ZeroModeResultPsi21})-(\ref{tildeGammaPhi}) for charged scalar decays.
Our methods is not suitable to compute the heavy neutrino momentum distribution  when the heavy neutrinos are produced in scatterings. It is, however, still possible to determine their total number density in such scenarios from (\ref{RateEquation}), (\ref{RateEquationPhi})  and by the methods described in Sec.~\ref{Sec:VollesRohr}.

If the scalar is a gauge singlet and decays as $\phi\rightarrow NN$, then the $N$-particles in the final state are not significantly affected by the plasma as long as their occupation numbers remain well below the equilibrium value, which is the case in many scenarios studied in the literature.
In this case the main thermal correction comes from the fact that $\phi$-quasiparticles in a thermal bath are screened and pick up a thermal mass.
The thermal mass correction appears even if $\phi$ is a gauge singlet because it must have some interactions to be produced in the early universe. The effective mass $M_\phi$ in the plasma grows with $T$ and the coupling constant $\lambda$; for $T\gg m_\phi/\sqrt{\lambda}$ it is much larger than the vacuum mass $m_\phi$. 
Since the decay rate is larger for heavier quasiparticles ($\tilde{\Gamma}_{\phi\q}\propto M_\phi$), this enhances the DM production rate significantly.
Equation (\ref{singletGammaArbitraryfN2}) gives the thermally corrected production rate $\tilde{\Gamma}_{\phi\q}$ in this case.
The effective mass $M_\phi$ therein can be determined once the $\phi$-interactions are specified. 
The momentum distribution of the DM can be calculated from (\ref{sigletmomentutmdependentproduction}).

If the scalar $\Phi$ is charged under some gauge group, then at least one of the final state particles in the decay (which we call $\Psi$) must also carry a charge. This particle is typically in equilibrium in the early universe. This leads to Pauli blocking as soon as the temperature exceeds the scalar mass ($T>m_\Phi$). The behaviour at higher temperatures depends on the effective masses $M_\Phi$ and $\fermionmass{M}_f$ of $\Phi$ and $\Psi$. If $M_\Phi\gg \fermionmass{M}_f$ at all temperatures, then the dominant contribution to $\tilde{\Gamma}_{\Phi\q}$ at all temperatures comes from the decay $\Phi\rightarrow \Psi N$, and the decay products with energies 
$\Upomega_{\Psi }\simeq \Upomega^+$
and 
$\Upomega_N \simeq \omega_N$ 
are always relativistic. In this case the rate $\tilde{\Gamma}_{\Phi\q}$ is given by (\ref{GammaqExpressionForDoublet}), as illustrated in Fig.~\ref{FullRateDoublet2}. 
The behaviour can easily be understood as the interplay between Pauli blocking, the enhancement of the decay rate due to the effective $\Phi$-mass $M_\Phi$ and the increased lifetime of a particle that moves in the plasma due to time dilatation (to be calculated with thermal masses): $\tilde{\Gamma}_{\Phi\q}$ is suppressed for $T\sim m_\Phi$ due to Pauli blocking, but grows linearly with $T$ for $M_\Phi\gg m_\Phi$. 
This situation is usually realised if $\Phi$ has some other interactions (in addition to the gauge coupling that increases the mass of both, $\Phi$ and $\Psi$).
We performed the calculation explicitly for an U(1) gauge interaction of $\Phi$. At the order in perturbation theory considered here, it is straightforward to obtain the results for each component of a general SU(N) multiplet from (\ref{ZeroModeResultPsi}), \eqref{GammaqExpressionForDoublet} and (\ref{QnonzeroResult}) by making the replacements described after (\ref{fermionmassdef}).
Hence, analytic results allow for a simple inclusion of thermal corrections in the regime where $N$ is predominantly produced in quasiparticle decays of U(1) or SU(N) charged scalars.
The DM momentum distribution can be obtained from (\ref{doubletmomentutmdependentproduction}).

If at some temperature $\fermionmass{M}_f$ becomes comparable to $M_\Phi$, then the $\Psi$-quasiparticles that are produced in the $\Phi\rightarrow \Psi N$ decay are not relativistic. In this case the decay rate becomes sensitive to the complicated infrared behaviour of fermion dispersion relations in a plasma. For $\fermionmass{M}_f>M_\Phi$ the decay may even become kinematically forbidden.
At the same time other processes, such as inelastic scatterings, contribute to heavy neutrino production. 
In this scenario it seems difficult to obtain an analytic expression for the scalar damping rate $\tilde{\Gamma}_{\Phi\q}$.
Qualitatively the behaviour is similar to (\ref{QnonzeroResult}), as illustrated in Fig.~\ref{FullRateDoublet}: $\tilde{\Gamma}_{\Phi\q}$ is suppressed for $T\sim m_\Phi$ due to Pauli blocking, but grows linearly with $T$ for $M_\Phi\gg m_\Phi$. However, we cannot easily determine the coefficient in the relation $\tilde{\Gamma}_{\Phi\q}\propto T$ in the high temperature regime. 
The precise value of this factor depends on the gauge group, the coupling constant and the thermodynamic state of $\Phi$, and it can only be determined numerically.

\section*{Acknowledgements}
We would like to thank Bj\"orn Garbrecht and Giovanni Villadoro for helpful discussions on the effect of scatterings and Landau damping and on the kinematics of particle decays, respectively.
This work was supported by the Gottfried Wilhelm Leibniz program of the Deutsche Forschungsgemeinschaft (DFG), the DFG cluster of excellence Origin and Structure of the Universe, the Jiangsu Ministry of Science and Technology under contract BK20131264 and Natural Science Foundation of China under contract 11405084,
the Projektbezogener Wissenschaftleraustausch program of the Bayerisches Hochschulzentrum f\"ur China and the visitor program of the Kavli Institute for Theoretical Physics China (KITPC) in Beijing. We also acknowledge the Priority Academic Program Development for Jiangsu Higher Education Institutions (PAPD).

\begin{appendix}

\section{Nonequilibrium quantum field theory}\label{methods}
\paragraph{Correlation functions in a medium} - The usual methods to calculate S-matrix elements are not suitable to describe nonequilibrium systems at large density because there is no well-defined notion of asymptotic states, and the properties of quasiparticles in a medium may significantly differ from those of particles in vacuum.
However, observables can always be expressed in terms of correlation functions of the quantum fields, without reference to asymptotic states or free particles. 
There are two independent two point functions for each field. 
For a real scalar field $\phi$ these are often chosen to be the connected Wightman functions  
\begin{eqnarray}
\Delta^>(x_1,x_2) = \langle \phi(x_1)\phi(x_2)\rangle_{c} \label{forwAA} \ , \ 
\Delta^<(x_1,x_2) = \langle \phi(x_2)\phi(x_1)\rangle_{c} \label{backAA}, 
\end{eqnarray}
where the $\langle\ldots\rangle$ is to be understood in the sense of the usual quantum statistical average $\langle\ldots\rangle={\rm Tr}(\varrho \ldots) $ of a system characterised by a density operator $\varrho$.
For a fermion $\Psi$ the corresponding definitions are
\begin{eqnarray}
S^{>}_{\alpha\beta}(x_{1},x_{2})=\langle \Psi_{\alpha}(x_{1})\bar{\Psi}_{\beta}(x_{2})\rangle_{c} \ , \
S^{<}_{\alpha\beta}(x_{1},x_{2})=-\langle \bar{\Psi}_{\beta}(x_{2})\Psi_{\alpha}(x_{1})\rangle_{c}.
\label{SbackA}
\end{eqnarray}
Here $\alpha$ and $\beta$ are spinor indices, which we suppress in the following.
We will mostly be concerned with fermions, and the Wightman functions $S^\gtrless$ will be needed to calculate Feynman diagrams.
Their linear combinations
\begin{eqnarray}
\Delta^{-}(x_{1},x_{2})&=&i\left(\Delta^>(x_1,x_2)-\Delta^<(x_1,x_2)\right)\label{dminus}\\
\Delta^{+}(x_{1},x_{2})&=&\frac{1}{2}\left(\Delta^>(x_1,x_2)+\Delta^<(x_1,x_2)\right)\label{dplus}\\
S^{-}(x_{1},x_{2})&=&i\left(S^{>}(x_{1},x_{2})-S^{<}(x_{1},x_{2})\right)\label{SMinus}\\
S^{+}(x_{1},x_{2})&=&\frac{1}{2}\left(S^{>}(x_{1},x_{2})+S^{<}(x_{1},x_{2})\right)\label{Splus}
\end{eqnarray}
have an intuitive physical interpretations. 
The \textit{spectral function} $S^{-}$ roughly speaking characterises the spectrum of quasiparticles in the plasma. 
If $\Psi$ is weakly coupled to a thermal bath, then $S^-(x_1,x_2)=S^-(x_1-x_2)$ depends only on the relative coordinate even if $\Psi$ itself is out of equilibrium and $S^+$ is time dependent \cite{Anisimov:2008dz}. Then we can express $S^-$  as the Fourier transform of the spectral density $\uprho(q)$,
\begin{equation}
S^-(x_1-x_2)=i\int \frac{d^4q}{(2\pi)^4}\uprho(q)\,e^{-i q (x_1-x_2)} ,
\end{equation}
which can be viewed as a definition for $\uprho(q)$.
In a weakly coupled theory, the pole structure of $\uprho(q)$ determines the dispersion relations in the plasma: The real parts of poles correspond to quasiparticle energies, the imaginary parts are the thermal widths of the quasiparticles, see (\ref{dispersionrelation}) and (\ref{GammaFormula}). 
The \textit{statistical propagator} $S^{+}$ provides a measure for the occupation numbers. 
The correlators fulfil the symmetry relations
\begin{eqnarray}
\gamma_{0}S^{-}(x_{2},x_{1})&=&-\left(\gamma_{0}S^{-}(x_{1},x_{2})\right)^{\dagger}\label{Sminussym}\\
\gamma_{0}S^{+}(x_{2},x_{1})&=&\left(\gamma_{0}S^{+}(x_{1},x_{2})\right)^{\dagger}\label{Splussym}
,\end{eqnarray}
which can be seen from the definitions.
If $\Psi$ is a Majorana fermion, then there is an additional symmetry
\begin{equation}\label{MajoranaSymmetry}
S^\gtrless(x_1,x_2)=C S^\gtrless(x_2,x_1)^T C^\dagger,
\end{equation}
where $C$ is the charge conjugation matrix.
$S^-$ has the boundary condition
\begin{eqnarray}
S^{-}(x_{1},x_{2})|_{t_{1}=t_{2}}&=&i\gamma_{0}\delta(\textbf{x}_{1}-\textbf{x}_{2})\label{DiracEqualTime}
,\end{eqnarray}
which follows from the canonical anticommutation relations. 
The boundary conditions for the statistical propagator are determined by the physical initial conditions in which the system is prepared.

\paragraph{Kadanoff-Baym ansatz} - 
It is reasonable to assume that all fields with gauge interactions are in thermal equilibrium at the time when $N$ gets produced. 
We will adopt that assumption in this work.
The scalar $\phi$ may or may not be in equilibrium, depending on whether $N$ production during freeze-in is relevant. 
Nonequilibrium propagators for scalars have e.g. been studied in \cite{Anisimov:2008dz,Drewes:2012qw,Garbrecht:2011xw,Millington:2012pf}. 
We will, however, not need them here because $\phi$ only appears as external particle. 
In thermal equilibrium at temperature $T$,\footnote{We work in the rest frame of the bath, where $T$ has a physical interpretation as temperature. Due to the choice of frame the expressions are not Lorentz-invariant. This choice is for convenience, the theory can of course be formulated in a covariant manner \cite{Weldon:1982aq}.}
the correlation functions only depend on the relative coordinate $x_{1}-x_{2}$,\footnote{
Vacuum can be seen as a special case of thermal equilibrium with $T=0$.
} 
and  they are related by the Kubo-Martin-Schwinger (KMS) relations, 
which are most conveniently written in terms of their Fourier transforms. In absence of chemical potentials they read
\begin{eqnarray}\label{KMS1}
\begin{tabular}{c c c}
$\Delta^{<}(q)=e^{-q_0/T}\Delta^{>}(q)$& , &$S^{<}(q)=-e^{-q_0/T}S^{>}(q)$
\end{tabular}
.\end{eqnarray}
These allow to express the momentum space two-point functions in terms of the spectral densities,
\begin{eqnarray}\label{KMSfermi}
\begin{tabular}{c c c}
$S^{-}(q)=i\uprho(q)$& , &$S^{+}(q)=\left(\frac{1}{2}-f_{F}(q_0)\right)\uprho(q)$,\\
$S^{>}(q)=(1-f_{F}(q_0))\uprho(q)$& , & $S^{<}(q)=-f_{F}(q_0)\uprho(q)$.
\end{tabular}
\end{eqnarray}
Here $f_{F}(q_0)=(e^{q_0/T}+1)^{-1}$ is Fermi-Dirac distribution, which naturally arises as a consequences of the boundary conditions for the correlation functions. 

The KMS-relations (\ref{KMSfermi}) in principle do not hold for $N$ because the sterile neutrinos are not in thermal equilibrium. Explicit expressions for the propagators of Majorana fermions out of thermal equilibrium have been found in \cite{Anisimov:2010aq,Anisimov:2010dk,Beneke:2010wd,Garny:2011hg,Dev:2014laa,Dev:2014wsa}, but lead to rather complicated calculations when being used in loop integrals.
Since $N$ couples weakly to $\phi$ and the thermal bath, the time scale $1/\Upgamma_{N\p}$ on which the occupation numbers evolve is slow compared to the time scale of microscopic processes in the plasma. This justifies to assume that a relation of the form
\begin{equation}\label{KBansatz}
S_N^{+}(q)=\left(\frac{1}{2}-f_N(q_0)\right)\uprho_N(q)
\end{equation}
holds locally, which is known as Kadanoff-Baym ansatz. Here $f_N$ is a function that can be interpreted as $N$-distribution function. It changes on macroscopic time scales $\gg 1/\Gamma_{\phi\q}$. 
It is known that this ansatz does not exactly hold \cite{Garbrecht:2011xw,Drewes:2012qw}, but it should be sufficient for the present purpose.
The Kadanoff-Baym ansatz allows to postulate relations similar to (\ref{KMSfermi}) also for $N$ locally in time, in spite of the fact that $N$ is not in equilibrium,
\begin{eqnarray}\label{KMSN}
\begin{tabular}{c c c}
$S_N^{-}(q)=i\uprho_N(q)$& , &$S_N^{+}(q)=\left(\frac{1}{2}-f_{N}(q_0)\right)\uprho_N(q)$,\\
$S^{>}(q)=(1-f_{N}(q_0))\uprho_N(q)$& , & $S^{<}(q)=-f_{N}(q_0)\uprho_N(q)$.
\end{tabular}
\end{eqnarray}
By using a single scalar function $f_N$, we already assumed that the different helicity states of $N$ all have the same occupation number, which seems reasonable.
The function $f_N$ must fulfil the relation
\begin{equation}\label{NegativefN}
f_N(-q_0)=1-f_N(q_0),
\end{equation}
which  follows directly from the Majorana condition (\ref{MajoranaSymmetry}) 
and does not rely on any assumption about the phase space distribution function of $N$-particles. 
This can alternatively be found by demanding that there is no charge associated with $N$ present in the plasma, which corresponds to the requirement that 
\begin{equation}
\int \frac{dq_0}{2\pi}{\rm tr}[\gamma_0S^+(q)]=0.
\end{equation}

If evaluated at on-shell values $q_0=\Upomega_{N \q}>0$, where $\Upomega_{N \q}$ is the mass shell of $N$-(quasi)particle, the function $f_N(\Upomega_{N \q})$ can be interpreted as the physical phase space distribution function $f_{N\q}$ of these quasiparticles in the plasma,
\begin{equation}
f_{N\q}=f_N(\Upomega_{N \q}) \ . \phantom{XXX} \  \label{PhaseSpaceDistrDef_App}
\end{equation}

\paragraph{Brownian motion} - 
If $\phi$ is weakly coupled to a thermal bath, then dissipative effects can be described by a damping rate $\Gamma_{\phi\q}$, which  can be extracted from the pole structure of the scalar spectral density 
\begin{equation}
\rho_\phi(q)=-i\Delta^-(q)=-i\int\frac{d^4 x}{(2\pi)^4} \Delta^-(x)\, e^{i q x}.\label{rhodefinition}
\end{equation}
For a scalar $\phi$ that couples to a plasma in equilibrium, $\rho_\phi(q)$ at leading order can be expressed as
\begin{eqnarray}\label{spectralfunction2}
\rho_\phi(q)={-2{\rm Im}\Pi_\phi^R(q)+2q_0\epsilon\over 
(q_0^2-m_\phi^2-\q^2-{\rm Re}\Pi_\phi^R(q))^2+({\rm Im}\Pi_\phi^R(q)+q_0\epsilon)^2}
\end{eqnarray}
even if $\phi$ itself is not in equilibrium  \cite{Anisimov:2008dz}.
Here the retarded self-energy $\Pi_\phi^R(q)$ depends on $T$.
Let $\hat{\Omega}_\q$ be a pole of $\rho_\phi(q)$, with 
\begin{eqnarray}\label{OmegaAndGammaDefs}
\Omega_{\phi \q}\equiv{\rm Re}\hat{\Omega}_\q \ {\rm  and} \ \Gamma_{\phi\q}\equiv 2{\rm Im}\hat{\Omega}_\q.
\end{eqnarray}
In weakly coupled theories one observes the hierarchy
\begin{equation}\label{quasiparticle}
\Gamma_{\phi\q}\ll\Omega_{\phi \q} 
\end{equation}
and can make the Breit-Wigner approximation
\begin{equation}\label{BWphi}
\rho_\phi^{\rm BW}(q)=2\mathcal{Z}\frac{q_0\Gamma_{\phi\q}}{(q_0^2-\Omega_{\phi\q}^2)^2+(q_0\Gamma_{\phi\q})^2} + \rho_\phi^{\rm cont}(q),
\end{equation}
with the residue
\begin{equation}\label{residue}
\mathcal{Z}=\left[1-\frac{1}{2\Omega_{\phi \q}}\frac{\partial {\rm Re}\Pi_\phi^R(q)}{\partial q_0}\right]^{-1}_{q_0=\Omega_{\phi \q}}
.\end{equation}
Then the dispersion relations can be obtained by solving the equation
\begin{equation}\label{dispersionrelation}
\Omega_{\phi \q}^2-\q^2-m_\phi^2-{\rm Re}\Pi_\phi^R(q)\big|_{q_0=\Omega_{\phi \q}}=0, 
\end{equation}
and the $\phi$-damping rate $\Gamma_{\phi\q}$ that we aim to calculate is given by the imaginary part of the retarded self energy at the quasiparticle pole $q_0=\Omega_{\phi \q}$,
\begin{equation}\label{GammaFormula}
\Gamma_{\phi\q}=-\mathcal{Z}\frac{{\rm Im}\Pi_\phi^R(q)}{q_0}\Big|_{q_0=\Omega_{\phi \q}}.
\end{equation}
$\Gamma_{\phi\q}$ and $\Omega_{\phi \q}$ depend on $T$ due to the temperature dependence of $\Pi_\phi^R(q)$.
The appearance of the imaginary part can be interpreted as finite temperature version of the optical theorem.
The rate (\ref{GammaFormula}) is the thermal damping rate for $\phi$-quasiparticles.
The contribution to it from Feynman diagrams involving $N$ can be used to determine production rate for $N$-particles. We have assumed that the screened single particle states are the only relevant quasiparticles, i.e., collective scalar excitations \cite{Drewes:2013bfa} play no role. 
If the $N$-particles are produced from the decay of a coherent condensate $\varphi\equiv\langle\phi\rangle$, the rate (\ref{GammaFormula}) is still applicable if this production happens during oscillations near the minimum of the effective potential $\mathcal{V}(\varphi)$, see e.g. \cite{Cheung:2015iqa}.
Analogous to $\Delta^\gtrless$, one can introduce self-energies $\Pi_\phi^\gtrless$\footnote{In the Schwinger-Keldysh-formalism, $\Pi_\phi^<(x_1,x_2)$ corresponds to a self-energy where the first time argument lies on the ``forward'' part of the closed time path and the second argument lies on the ``backwards'' part. For $\Pi_\phi^>(x_1,x_2)$ it is the other way around.} and define
\begin{eqnarray}
\Pi_\phi^-(q)\equiv\Pi_\phi^>(q)-\Pi_\phi^<(q)=2i {\rm Im}\Pi_\phi^R(q). \label{UsefulRelation1}
\end{eqnarray}
This allows to rewrite 
\begin{equation}
\Gamma_{\phi\q}
=\mathcal{Z}\frac{i\Pi_\phi^-(q)}{2q_0}\Big|_{q_0=\Omega_{\phi \q}}\label{GammaFormula2}
=\Gamma_{\phi\q}^>-\Gamma_{\phi\q}^<,
\end{equation}
where 
\begin{eqnarray}
\Gamma_{\phi\q}^<&=&\mathcal{Z}\frac{i\Pi_\phi^<(q)}{2\Omega_{\phi \q}}|_{q_0=\Omega_{\phi \q}}\label{gain}\\
\Gamma_{\phi\q}^>&=&\mathcal{Z}\frac{i\Pi_\phi^>(q)}{2\Omega_{\phi \q}}|_{q_0=\Omega_{\phi \q}}\label{loss}
\end{eqnarray}
are the gain and loss rate for $\phi$-particles, which in total give the thermal damping rate $\Gamma_{\phi\q}$.
The thermal width $\Gamma_{\phi\q}$ does not only determine the damping of the spectral function $\Delta^-$ via (\ref{spectralfunction2}), it also governs the relaxation of the statistical propagator to its equilibrium form, which in the approximation (\ref{quasiparticle}) reads \cite{Drewes:2010pf} 
\begin{eqnarray}
\lefteqn{\Delta^{+}_\q(t_1,t_2)\simeq
\frac{\Delta^{+}_{\q;\text{in}}}{2}
\left[\cos(\Omega_{\phi\q}(t_1+t_2)
+\cos[\Omega_{\phi\q}(\tm)]\right]e^{-\Gamma_{\phi\q}\tp}}\nonumber\\
&+&\frac{\dot{\Delta}^{+}_{\q;\text{in}}}{\Omega_{\phi\q}}
\sin[\Omega_{\phi\q}(t_1+t_2)]e^{-\Gamma_{\phi\q}\tp}-\frac{\ddot{\Delta}^{+}_{\q;\text{in}}}{2\Omega_{\phi\q}^{2}}
\left[\cos[\Omega_{\phi\q}(t_1+t_2)]-\cos[\Omega_{\phi\q}(\tm)]\right)e^{-\Gamma_{\phi\q}\tp}\nonumber\\
&+&\frac{\coth(\frac{\Omega_{\phi\q}}{2T})}{2\Omega_{\phi\q}}\cos[\Omega_{\phi\q}(\tm)]\left(1-e^{-\Gamma_{\phi\q}\tp}\right)e^{-\nicefrac{\Gamma_{\phi\q}|\tm|}{2}},\label{narrowPlus}
\end{eqnarray}
where $\Delta^{+}_{\q;\text{in}}$, $\dot{\Delta}^{+}_{\q;\text{in}}$ and $\ddot{\Delta}^{+}_{\q;\text{in}}$ are boundary conditions for the statistical propagator and its derivatives at $t_1=t_2=0$. The choice 
    $\Delta^{+}_{\q;\text{in}}
    = 
\frac{1}{\Omega_{\phi \q}}[\frac{1}{2}+f_{\phi\q}(0)]$, 
$\dot{\Delta}^{+}_{\q;\text{in}}
    = 0 $, 
$\ddot{\Delta}^{+}_{\q;\text{in}}
   =
\Omega_{\phi \q}[\frac{1}{2}+   f_{\phi\q}(0)])$, which can be interpreted as a quantum state with well-defined quasiparticle occupation number $f_{\phi\q}$, simplifies (\ref{narrowPlus})  to 
\begin{eqnarray}
\Delta^{+}_\q(t_1,t_2)\simeq\frac{\cos[\Omega_{\phi \q}(t_1-t_2)]}{\Omega_{\phi \q}}e^{-\Gamma_{\phi\q}(t_1-t_2)}\left[\frac{1}{2} + f_{\phi\q}(t)\right],
\end{eqnarray}
where $t=(t_1+t_2)/2$ and $f_{\phi\q}(t)=f_B(\Omega_{\phi \q})+[f_{\phi\q}(0)-f_B(\Omega_{\phi \q})]e^{-\Gamma_{\phi\q} t}$ is the solution to the Boltzmann equation
\begin{equation}\label{BEforf}
\partial_t f_{\phi\q}(t)=-\Gamma_{\phi\q}[f_{\phi\q}(t)-f_B(\Omega_{\phi \q})].
\end{equation}
Here $f_B(\Omega)= (e^{\Omega/T}-1)^{-1}$ is the Bose-Einstein distribution. Finally,  the modes of the mean field $\varphi=\langle\phi\rangle$ obey an equation of motion of the form
\begin{eqnarray}
\ddot{\varphi}_\q+\Gamma_{\varphi\q}\dot{\varphi}_\q 
+ \partial_\varphi \mathcal{V}(\varphi_\q)
= 0,\label{FieldEqOfMot}
\end{eqnarray}
where $\mathcal{V}(\varphi)$ is the finite temperature effective potential.
Near its minimum $\Gamma_{\varphi\q}=\Gamma_{\phi\q}$ \cite{Cheung:2015iqa}. Hence, $\Gamma_{\phi\q}$ in this situation is the sole damping scale in the system \cite{Garbrecht:2011xw}.
With the KMS-relation $\Pi_\phi^<(q)=e^{-q_0/T}\Pi_\phi^>(q)$ it is easy to see that the detailed balanced relation
\begin{equation}\label{DetailedBalance}
\frac{\Gamma^<_{\phi\q}}{\Gamma_{\phi\q}^>}=e^{-\Omega_{\phi \q}/T} 
\end{equation}
holds and the total relaxation rate $\Gamma_{\phi\q}$ can be related to the $\phi$-production rate $\Gamma^<_{\phi\q}$ via \cite{Weldon:1983jn}
\begin{equation}
\Gamma^<_{\phi\q} = f_B(\Omega_{\phi \q})\Gamma_{\phi\q}.
\end{equation} 
These relations hold to all orders in the couplings amongst the thermal bath's constituents \cite{Bodeker:2015exa}.
Analogous to (\ref{gain}) and (\ref{loss}) one can find a damping rate $\Upgamma_\q$ for fermionic correlation functions $S^\pm$,
\begin{equation}
\Upgamma_\q^\gtrless=\mathcal{Z}_\q\frac{i{\rm tr}[\Slash{q}\,\Sigma^\gtrless(q)]}{2q_0}\big|_{q_0=\Upomega_{\Psi \q}},\label{FermionGamma}
\end{equation} 
where $\Upomega_{\Psi \q}$ is a fermionic quasiparticle pole and $\Sigma^\gtrless$ are fermionic self-energies define analogous to $\Pi_\phi^\gtrless$.

\paragraph{Damping rate far from equilibrium} - 
The expressions following (\ref{rhodefinition}) apply if all propagators Feynman diagrams that contribute to the self energies $\Pi_\phi^\gtrless$ fulfil KMS-relations like (\ref{KMS1}).
This is in good approximation the case if $\phi$ couples weakly to a larger thermal bath. 
This is not the case in the present calculation because neither the scalar field not $N$ are in equilibrium in general.
Without this assumption, the nonequilibrium correlation functions can be expressed as \cite{Drewes:2012qw}
\begin{eqnarray}
  \Delta^-_\q(t_1,t_2) &=& \frac{
  \sin\left(\int_{t_2}^{t_1}\!dt' \Omega_{\phi \q}(t')\right)
  e^{-\frac12\left|\int_{t_2}^{t_1}dt' \Gamma_{\phi\q}(t')\right|}
  }{2\sqrt{\Omega_{\phi \q}(t_1)\Omega_{\phi \q}(t_2)}},
  \label{eqn:Dm2}\\
  \Delta^+_\q(t_1,t_2) &=& \frac{
  \cos\left(\int_{t_2}^{t_1}\!dt' \Omega_{\phi \q}(t')\right)
  e^{-\frac12\left|\int_{t_2}^{t_1}dt' \Gamma_{\phi\q}(t')\right|}
  }{2\sqrt{\Omega_{\phi \q}(t_1)\Omega_{\phi \q}(t_2)}} \left[1 + 2f_{\phi\q}(t) \right],
  \label{eqn:Dp2}
\end{eqnarray}
where $t={\rm min}(t_1,t_2)$.
Note that (\ref{eqn:Dm2}) and (\ref{eqn:Dp2}) can be derived from first principles, without the Kadanoff-Baym ansatz (\ref{KBansatz}).
Here $f_{\phi\q}$ can be interpreted as the quantum mechanical generalisation of a phase space distribution function.
On macroscopic time scales, and if the damping is primarily driven by decays and inverse decays, its evolution is governed by the Markovian equation of motion 
\begin{equation}\label{EffBE}
  \partial_{t} f_{\phi\q}(t) = \left( 1+f_{\phi\q}(t) \right)\Gamma_{\phi\q}^<(t) - f_{\phi\q}(t)\Gamma_{\phi\q}^>(t).
\end{equation}
Formally this equation can be casted in the same form as (\ref{BEforf})
\begin{equation}
\partial_t f_{\phi\q}(t) = -\Gamma_{\phi\q}(t)\left[f_{\phi\q}(t)-\bar f_{\phi\q}(t)\right],\label{effBEsimple}
\end{equation} 
by defining
\begin{equation}
\bar f_{\phi\q}(t)\equiv(\Gamma_{\phi\q}^>(t)/\Gamma_{\phi\q}^<(t)-1)^{-1}.
\end{equation}
In the special case that $\phi$ couples to a large thermal bath, one can easily recover (\ref{BEforf}) from (\ref{effBEsimple}) by using (\ref{DetailedBalance}). In this case $\bar{f}_{\phi\q}\rightarrow f_B$ is time independent, and its values for all modes $\q$ are fixed by a single parameter $T$.
In general this is, however, not the case.

\end{appendix}

\bibliographystyle{JHEP}
\bibliography{all}

\end{document}